%% file: arxiv_main.tex
\RequirePackage[T1]{fontenc}
\documentclass[journal,10pt,letterpaper]{IEEEtran}
\usepackage{amsmath,amssymb,amsthm}
\usepackage{graphicx,booktabs,array}
\usepackage{cite}
\usepackage{algorithm}
\usepackage{algpseudocode}
\usepackage[hidelinks]{hyperref}
\newtheorem{theorem}{Theorem}
\newtheorem{lemma}{Lemma}
\newtheorem{proposition}{Proposition}
\newtheorem{corollary}{Corollary}
\newtheorem{assumption}{Assumption}
\newcommand{\E}{\mathbb E}
\newcommand{\R}{\operatorname{Re}}
\newcommand{\Hm}{\operatorname{Herm}}
\newcommand{\tr}{\operatorname{tr}}
\newcommand{\norm}[1]{\left\lVert#1\right\rVert}
\newcommand{\F}{\mathcal F}
\newcommand{\Lc}{\mathcal L}
\newcommand{\mhat}{\widehat{\boldsymbol m}}
\usepackage{enumitem}
\theoremstyle{definition}

\theoremstyle{remark}

\theoremstyle{plain}
\allowdisplaybreaks
\begin{document}
\bstctlcite{IEEEexample:BSTcontrol}
\setlength{\abovedisplayskip}{5pt plus 2pt minus 1pt}
\setlength{\belowdisplayskip}{5pt plus 2pt minus 1pt}
\setlength{\abovedisplayshortskip}{0pt plus 2pt}
\setlength{\belowdisplayshortskip}{1.5pt plus 2pt minus 0.5pt}
\title{Keep Pace to Learn: Movable-Antenna Compensation of Channel Aging in Over-the-Air Split Learning}
\author{Xianzhong~Li, Yuzhi~Yang, Tianhao~Guo, Zhaoyang Zhang
\vspace{-0.8cm}%
\thanks{Xianzhong Li and Tianhao Guo are with the College of Physics and Electronic Engineering, Shanxi University, Taiyuan 030006, China (e-mail: \{lixianzhong, tianhao\_guo\}@sxu.edu.cn).}%
\thanks{Yuzhi Yang is with the College of Information Science, Japan Advanced Institute of Science and Technology, Nomi 9231292, Japan (e-mail: yangyzis@jaist.ac.jp).}%
\thanks{Zhaoyang Zhang is with the College of Information Science and Electronic Engineering, Zhejiang University, and also with Zhejiang Provincial Key Laboratory of Information Processing, Communication and Networking (IPCAN), Hangzhou 310027, China (e-mail: zhzy@zju.edu.cn).}
\thanks{Corresponding author: Tianhao Guo.}\thanks{This is an extended version of a manuscript submitted to the IEEE Journal on Selected Areas in Communications. Appendices~C--K contain material that the submitted version omits for space.}}
\maketitle
\begin{abstract}
Split learning (SL) trains a neural network across devices by exchanging intermediate activations and gradients. In over-the-air SL, the wireless channel serves as an inter-layer linear map, and when the forward and backward channels match, reciprocity and conjugation supply the backward product without channel estimation at the base station (BS). Passive user motion, which the learning link does not control, breaks this match. Movable antennas let the antenna move relative to the user device, and we use this freedom to compensate the displacement. For a stationary BS and a propagation environment quasi-static within each exchange, returning the antenna to its forward-transmission position restores the channel and yields a conditionally unbiased gradient estimate. We also derive when discrete positioning or mechanical movement can achieve this in time. A residual displacement below a quarter wavelength keeps the real part of the channel correlation positive for any angular spectrum. At zero effective offset, independent Gaussian jitter attenuates this correlation without reversing its sign, whereas a persistent offset can reverse it, and in arrays the tolerance depends on the receive-combining direction. These effects enter a nonconvex convergence bound, and a local analysis shows that partial mechanical compensation saves cycle time to first order at only a second-order loss in correlation. In CIFAR-10 experiments, a user terminal without channel state information stays at chance level under uncompensated walking motion, while discrete positioning stays within about 0.3 percentage points of exact restoration, including at a speed beyond the modeled actuator's limit.
\end{abstract}
\section{Introduction}
Future wireless systems are expected to integrate communication, computation, and intelligence, so that neural network (NN) components deployed on different devices are trained jointly over wireless links. Split learning (SL) trains such an NN across devices by exchanging intermediate activations and their gradients~\cite{gupta2018,wu2023sl}. In over-the-air SL, the wireless multiple-input multiple-output (MIMO) channel can serve as the linear connection between these components~\cite{yang2023}. The base station (BS) sends activations on the downlink, and the user equipment (UE) computes the loss and returns the cut-layer gradient, that is, the loss gradient with respect to the received signal. When the two transmissions see the same calibrated channel, conjugating the gradient before the uplink transmission and the received signal at the BS applies the conjugate-transposed downlink channel to the gradient. This is exactly the backward product that backpropagation requires, obtained without channel estimation at the BS.

The backward product is exact when the channel during gradient feedback equals the channel used in the forward computation. Reciprocity, together with a propagation field that stays static within the exchange, gives this equality at a fixed position, not across positions: if the UE moves before feedback, the forward signal may remain useful, but the gradient is multiplied by a different channel matrix.

A handheld or wearable UE can move during transmission, local computation, and feedback scheduling. This passive motion is imposed by the user rather than the learning link~\cite{wei2025passive,shih2025localization}. At $28$~GHz, a user moving at $1$~m/s travels about half a wavelength during a $5$~ms forward--backward interval. The resulting path-phase changes can attenuate or reverse the useful descent component of the returned gradient.

Movable antennas (MAs) allow the antenna position to be adjusted relative to the UE body~\cite{zhu2024model}. We use this spatial freedom to compensate for passive UE movement. For a stationary BS and a propagation field quasi-static within each exchange, returning the calibrated antenna, before gradient feedback, to the world position at which it received the forward signal reproduces the forward channel. We call this position the target position. It is computed from the UE displacement and the exchange timing, so no search for a stronger channel realization is needed. Discrete positioning switches the active antenna to a calibrated trailing element, while a mechanical actuator moves the active antenna continuously against the UE motion. Their usefulness depends on aperture, motion limits, and whether the adjustment and any reset fit repeated exchanges.

Exact restoration is not always attainable or worth its cost. The target position may fall between the available antenna positions or require a compensation speed beyond the actuator limit, and a longer mechanical stroke lengthens the reset cycle. Even with ideal hardware, the target itself is an estimate. The UE computes it from its estimated speed and the scheduled forward--backward interval, so errors in motion estimation and timing, together with actuator positioning error, generally leave some residual. A persistent velocity bias shifts every exchange by the same offset, whereas random estimation and actuation errors scatter around the target, and these two kinds of error affect learning differently. Learning can nevertheless tolerate some residual mismatch. The tolerable residual depends on the receiver, the positioning-error distribution, and the time available between updates. Reducing a small residual improves gradient alignment but requires more mechanical travel and a longer reset time. Each exchange uses the UE's current location as a new forward reference, so compensation covers the displacement within one exchange rather than the distance traveled over training.

We first establish when antenna positioning can restore the forward channel and its backward product under the available hardware and timing limits. When a residual remains, bounds valid for any angular spectrum give a residual displacement below which the real part of the correlation stays positive, while separate offset and jitter analyses show why positioning errors with the same root-mean-square displacement need not have equal effects. With multiple receive antennas, paths arriving from different angles undergo different motion-induced phase rotations, and paths nearly perpendicular to the motion barely change phase. We therefore bound the alignment separately for each receive-combining direction, that is, each combination of antenna outputs, which a single correlation value cannot capture. Scalar and resolved-path models then connect these alignment properties to nonconvex learning, including the effect of power constraints on the last linear layer under path-mode precoding. Finally, leaving a small residual displacement shortens each mechanical cycle, and locally this first-order time saving outweighs the second-order loss in the real part of the correlation.

Experiments on CIFAR-10 and on a synthetic binary classification task test these predictions. Without compensation at walking speed, a UE without channel state information (CSI) remains at chance level, whereas an eight-antenna UE with a minimum mean-square-error (MMSE) receiver still learns. Removing the positive eigenspace of the Hermitian part of its mean combiner-output operator collapses this accuracy, whereas random-subspace removal and energy-matched controls do not. In the synthetic task, accuracy tracks the real part of the correlation rather than its magnitude, including a recovery at a residual beyond one wavelength, and a fixed bias reduces accuracy to chance level, whereas Gaussian jitter with the same root-mean-square displacement does not. Restoration matches training with an ideal backward product without the pilot overhead of digital correction, and in modeled exchange-and-reset time, discrete positioning reaches $90\%$ accuracy before the fixed and exact-mechanical implementations.

The main contributions of this paper can be summarized as follows.
\begin{itemize}
\item We establish channel-restoration and hardware-feasibility conditions for movable-antenna-assisted over-the-air SL, realized through discrete positioning or repeated mechanical compensation.
\item We derive residual-dependent alignment and nonconvex convergence conditions, distinguishing systematic offset from Gaussian jitter and revealing direction-dependent tolerance in arrays, confirmed by receiver-dependent CIFAR-10 results.
\item We characterize the alignment--cycle-time trade-off and demonstrate that discrete positioning can outperform exact mechanical restoration in modeled time to target accuracy.
\end{itemize}
\begin{figure*}[!htbp]
\centering\includegraphics[width=0.85\textwidth]{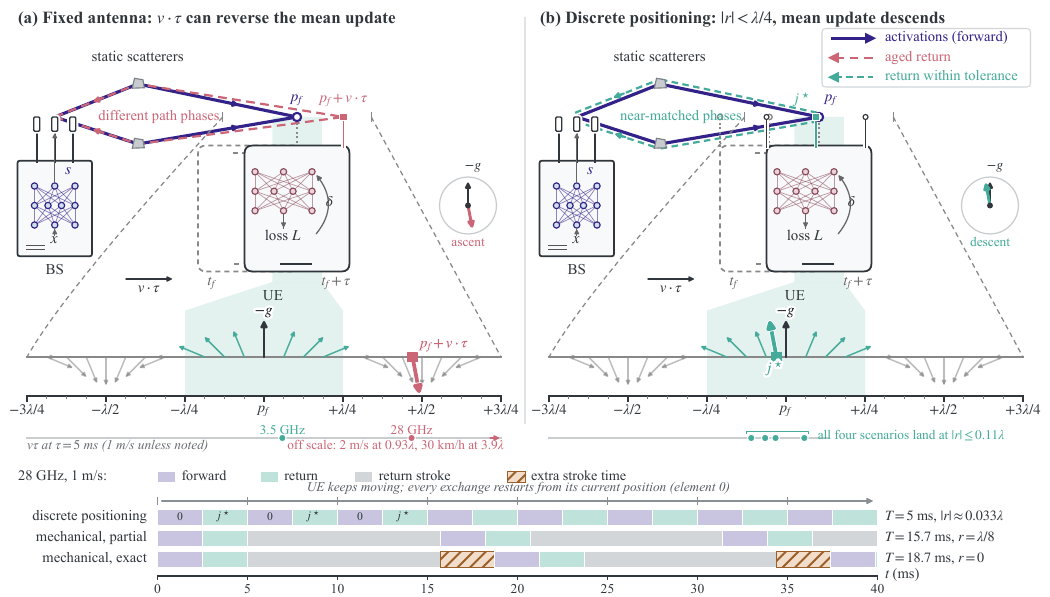}\par
\caption{Over-the-air SL under passive UE motion, with forward activations from the BS and a cut-layer gradient returned by the UE. The strips illustrate scalar conditional-mean alignment: black arrows denote $-g$ and colored arrows the corresponding mean updates. Shading marks $|r|<\lambda/4$, and the compasses compare the realized mean update with $-g$. (a) A fixed antenna returns feedback from $p_f+v\tau$, where the mean update can be an ascent direction. (b) Discrete positioning switches to the element $j^\star$ nearest $p_f$. Phases match exactly only at zero residual. Markers show four $v\tau$ examples at $\tau=5$~ms. The bottom shows the modeled $28$~GHz, $1$~m/s exchange cadence, with a new reference each exchange.}
\label{fig:system}
\end{figure*}

\section{Related Works}
\subsection{Over-the-Air Learning}
Over-the-air computation exploits wireless superposition to combine information from distributed devices. Broadband analog aggregation builds on it to reduce the communication cost of federated edge learning~\cite{zhu2020baa}, and over-the-air federated learning and distributed stochastic gradient descent (SGD) integrate such aggregation into model optimization~\cite{yang2020otafl,amiri2020,abyaneh2025otafl}. Extensions consider scalable cell-free deployments~\cite{sifaou2024cellfree} and aggregation without explicit channel estimation~\cite{zhong2024estfree}. Movable antennas have also been used to reshape the aggregation channel~\cite{li2025aircompma,mao2026marisaircomp}. In these settings, the wireless operation combines local updates, so channel distortion affects the aggregate supplied to the optimizer.

SL places the communication interface inside the neural network. Devices exchange intermediate activations and their gradients instead of complete model updates~\cite{gupta2018}. Wireless SL consequently requires a joint design of model partitioning, parallel execution, and radio resource management~\cite{wu2023sl,xu2024splitfl,zhou2026adaptivesfl}. Mobility-aware split federated learning extends this setting to semantic communication among mobile users~\cite{zheng2024mobilitysfl}. SL also supports semantic alignment among neural transceivers by adapting their encoder and decoder components~\cite{choi2023semantics}. Related research studies neural-network transmission and physical neural links over wireless channels~\cite{jankowski2024airnet,cang2026neurallink}. These directions motivate tighter integration between communication and neural processing, with different requirements for transmitting models, features, and training signals.

Yang \emph{et al.}~\cite{yang2023} embed the MIMO channel in an inter-layer connection and use channel reciprocity and conjugation to implement backpropagation through the channel. Their experiments allow channels to evolve across exchanges, but each backward product corresponds to a matched forward operator. Channel-aging and delayed-feedback studies characterize communication losses~\cite{truong2013,au2008mimosvd,ding2026hsraging}, while prediction methods estimate later channel states~\cite{yin2020,wang2024prediction}. Companion physical-backpropagation studies correct gradient feedback over an aged channel digitally or delay it to exploit favorable spatial correlation~\cite{twc,wcl}. Here, antenna-position control compensates the spatial mismatch before the backward transmission, subject to finite positioning and timing resources.
\subsection{Movable Antennas}
Movable antennas adjust antenna coordinates to exploit spatial variation in the propagation field. Field-response models describe how the phases of multiple paths change with position and provide a basis for communication-performance optimization~\cite{zhu2024model}. Overviews and tutorials survey the opportunities, hardware architectures, and implementation challenges of this spatial control~\cite{zhu2024overview,zhu2025tutorial,ning2025architectures}. Position optimization can use field-response information obtained through channel estimation~\cite{jang2025maestimation}, statistical CSI~\cite{chen2024statcsi}, or separate timescales for antenna movement and communication processing~\cite{hu2025twotimescale}. Six-dimensional designs additionally optimize orientation~\cite{shao2025sixdma,shao2025sixdmatutorial}, and recent work examines reconfiguration under high mobility~\cite{ji2026iov6dma}.

Implementation determines how quickly a desired position can be realized. Prototype studies demonstrate physical antenna positioning~\cite{dong2024prototype}, while electronic movable-antenna arrays provide reconfiguration through electronic control~\cite{chen2025remaa}. Pinching-antenna systems reconfigure radiation points along waveguides and provide another architecture for spatially adjustable transmission~\cite{ding2025pinching,ouyang2025passgain,fang2025pinchingai}. Mechanical movement must account for travel, acceleration, reset, and motion power~\cite{wu2025globalma}. Electronic reconfiguration avoids a mechanical trajectory but requires available calibrated configurations. These architectures motivate discrete positioning for short exchanges and mechanical positioning when the device and schedule support repeated motion. They do not make a particular displacement automatically reachable within a prescribed feedback interval.

Spatial control has already been used to compensate terminal movement. Predictor antennas reuse a leading antenna's observation when a trailing antenna reaches the same location, with velocity-aware switching adapting this reuse~\cite{sternad2012,guo2021}. Counter-movement can preserve an antenna's world position~\cite{artner2020}, and antenna mobility has been studied for transforming time-varying channels into static ones~\cite{li2025static}. Passive movement can instead create virtual arrays for sensing and localization~\cite{wei2025passive,shih2025localization}. Here it is the source of forward--backward mismatch. The additional question is how much spatial mismatch learning can tolerate, and how that tolerance changes feasible positioning and repeatable update time.
\subsection{Learning with Imperfect Gradients}
Physical backpropagation conveys a gradient through an uncertain operator. Related adaptive-control studies show that errors in the modeled propagation or plant response can alter an update direction~\cite{morgan1980,boucher1991}. Multichannel analysis further distinguishes scalar phase uncertainty from matrix-valued distortion~\cite{elliott1992}. This distinction matters for antenna arrays because a mean channel correlation need not describe every spatial direction. Receive processing can also depend on the forward channel, so an unconditional operator average does not automatically characterize the gradient produced by an adaptive receiver.

Stochastic-optimization analysis accommodates imperfect and biased gradients through conditions on their mean alignment and moments~\cite{bottou2018}. Decentralized SGD with biased gradients provides another setting in which systematic gradient error enters convergence analysis~\cite{jiang2025biasedsgd}. These results motivate separating persistent positioning offset from random jitter and distinguishing correlation from a conditional descent guarantee. Our analysis derives such quantities from residual antenna displacement and states the extra conditions needed for scalar models, fixed combiners, adaptive processing, and power-constrained updates. The resulting guarantees connect a physically feasible antenna adjustment to optimization behavior, while the experiments assess learning performance under the specified channel and receiver models.

\section{System Model}\label{sec:model}
Fig.~\ref{fig:system} shows a stationary BS with $N_t$ fixed antennas executing the early network layers and transmitting their activations on the downlink. The UE has $N_r$ antennas adjustable relative to the UE body, executes the remaining layers, and computes the loss. It returns the cut-layer gradient on the uplink, allowing the BS to update its layers by reciprocal physical backpropagation~\cite{yang2023}.
\subsection{Communication Model}
For a given UE array reference position $\boldsymbol p$ in world coordinates, let $H(\boldsymbol p)\in\mathbb C^{N_r\times N_t}$ denote the downlink spatial channel from the stationary BS. We consider calibrated antenna responses, static scatterers, and negligible motion within each radio-frequency (RF) transmission. The forward and backward transmissions of an exchange share the same spatial channel function. Its value changes as the UE antenna samples different positions. 

Within a local region where far-field path directions and gains are approximately constant, the channel comprises $L_p$ propagation paths~\cite{zhu2024model}:
\begin{equation}\label{eq:multipath}
 H(\boldsymbol p)=\sum_{l=1}^{L_p}
 \alpha_l e^{j\kappa\boldsymbol u_l^T\boldsymbol p}
 \boldsymbol a_{r,l}\boldsymbol a_{t,l}^{T},
 \qquad \kappa=\frac{2\pi}{\lambda}.
\end{equation}

Here $\lambda$ is the carrier wavelength, $\alpha_l$ is the complex gain of path $l$ with its reference phase absorbed at the world-coordinate origin, and $\boldsymbol u_l$ is its unit arrival-direction vector at the UE. The vectors $\boldsymbol a_{t,l}\in\mathbb C^{N_t}$ and $\boldsymbol a_{r,l}\in\mathbb C^{N_r}$ are the fixed BS array-response vector and the UE subarray-response vector relative to its reference position, respectively. These path parameters remain constant within the modeled exchange and local region but may change between exchanges. With reciprocity calibration, the corresponding uplink operator at the same position is $H(\boldsymbol p)^T$.
\subsection{UE Motion Model}
Let $t_f$ and $t_b=t_f+\tau$ denote the forward and backward transmission times, respectively, with $\tau>0$. Subscripts $f$ and $b$ identify the forward and backward transmissions. The interval $\tau$ accounts for the elapsed time between the channel samples, including intervening transmission, UE computation, and feedback scheduling. It describes one exchange rather than the complete training duration.

We approximate passive UE motion during this interval by translation at constant speed $v\ge0$ in direction $\mhat$, with negligible rotation. This motion is imposed by the user rather than commanded by the communication link~\cite{wei2025passive,shih2025localization}. Let $\boldsymbol c_{\rm UE}(t)$ be the UE body position in world coordinates. The body displacement over the exchange is $v\tau\mhat$.
The BS position remains fixed. Constant velocity is a local approximation for one exchange and need not hold throughout training. The displacement is not restricted to a fraction of a wavelength. Its magnitude determines the adjustment required at the UE. Rotation, which six-dimensional movable antennas can address~\cite{shao2025sixdma,shao2025sixdmatutorial}, is outside the present model.
\subsection{Antenna Adjustment Model}
Let $\boldsymbol a_{\rm UE}(t)$ denote the active antenna or subarray reference position relative to the UE body. Its world position is $\boldsymbol p_{\rm UE}(t) =\boldsymbol c_{\rm UE}(t)+\boldsymbol a_{\rm UE}(t)$.
For a multi-antenna UE, adjustment translates the active subarray rigidly while preserving its internal geometry and orientation. A commanded displacement $q\ge0$ opposite to the body motion gives $\boldsymbol a_{\rm UE}(t_b)-\boldsymbol a_{\rm UE}(t_f)=-q\mhat$.
This displacement is realized by switching to a calibrated trailing element or subarray, or by mechanically moving the active antenna relative to the UE body. It changes the antenna offset without controlling the UE's passive motion.

Define $\boldsymbol p_f=\boldsymbol p_{\rm UE}(t_f)$ and $\boldsymbol p_b=\boldsymbol p_{\rm UE}(t_b)$. Combining the passive and controlled displacements yields
\begin{equation}\label{eq:geometry}
 \begin{aligned}
 \boldsymbol r
 &=\boldsymbol p_b-\boldsymbol p_f
 \\
 &=(v\tau-q)\mhat,
 \qquad r=v\tau-q.
 \end{aligned}
\end{equation}

The scalar $r$ is the signed residual displacement. With $q=0$, the antenna moves with the UE body. With $q=v\tau$, its backward-transmission position coincides with its forward position. This geometric relation specifies the target position. Whether it can be realized depends on the available antenna positions, mechanical travel, and exchange timing.
\subsection{Channel and Gradient Mismatch}
The two sampled channels are $H_f=H(\boldsymbol p_f),\qquad H_b=H(\boldsymbol p_b)$.
Both matrices use the downlink convention, so the physical uplink channel at the backward transmission time is $H_b^T$. Substituting $\boldsymbol p_b=\boldsymbol p_f+\boldsymbol r$ into the multipath model gives
\begin{equation}\label{eq:aged-multipath}
 H_b=\sum_{l=1}^{L_p}\alpha_l
 e^{j\kappa\boldsymbol u_l^T\boldsymbol p_f}
 e^{j\kappa\boldsymbol u_l^T\boldsymbol r}
 \boldsymbol a_{r,l}\boldsymbol a_{t,l}^{T}.
\end{equation}

Each path acquires a phase change determined by its arrival direction and the residual displacement. Thus $H_b$ can differ from $H_f$ even when electromagnetic reciprocity holds. The mismatch arises because the forward and backward transmissions sample different positions within the same propagation field.

The BS model maps input $x$ to $s=f_W(x)\in\mathbb C^{N_t}$, with trainable parameters $W$ and $\norm{s}\le1$. The UE receives the downlink signal and differentiates its real-valued loss $\Lc$:
\begin{equation}
 y=H_f s+n_f,\qquad
 \delta=2\frac{\partial\Lc}{\partial y^*}\in\mathbb C^{N_r}.
\end{equation}

Here $n_f\in\mathbb C^{N_r}$ is forward receiver noise, and $\delta$ includes differentiation through the UE-side processing. The UE transmits a power-normalized version of this gradient at $t_b$. After reciprocity calibration~\cite{xu2024reciprocity}, conjugation before transmission and after reception, and compensation for the known normalization, the BS obtains
\begin{equation}\label{eq:backward}
 \hat g=P(H_b^H\delta+w),\qquad g=P(H_f^H\delta).
\end{equation}

Here $w\in\mathbb C^{N_t}$ is backward noise and $P$ is the real-linear adjoint of the Jacobian of $f_W$, so $g$ is the parameter gradient required by the forward computation and $\hat g$ its physically received estimate.

The discrepancy between the backpropagated gradient estimate and the gradient required by the forward computation is
\begin{equation}\label{eq:gradient-mismatch}
 \hat g-g=P\bigl((H_b-H_f)^H\delta+w\bigr).
\end{equation}

Without adjustment, the antenna is a fixed-position antenna relative to the UE body, so it moves with the UE and samples the backward channel at a different position. Antenna control can remove this mismatch exactly or reduce the residual enough to preserve useful learning. The BS requires no channel estimate for the backward product. The UE may use forward CSI, and the experiments cover both cases.
\section{Motion-Compensating Antenna Control}\label{sec:speed}
\subsection{Residual Displacement and Positioning Uncertainty}

Motion estimation and antenna actuation need not be exact. The realized residual UE antenna displacement is $\boldsymbol r_{\rm act}=\boldsymbol b_{\rm eff}+\boldsymbol e$, where $\boldsymbol b_{\rm eff}=r\mhat+\boldsymbol b$.
Here $\boldsymbol b$ is a persistent motion-estimation or position-calibration offset and $\boldsymbol e$ an independent zero-mean positioning error, with covariance $\Sigma=\sigma^2\mhat\mhat^T$ for a rail and $\sigma^2I$ for an isotropic error.

To characterize spatial sensitivity, consider uncorrelated zero-mean path gains with powers $p_l=\E|\alpha_l|^2$. Let $\mu$ be their normalized angular power distribution, assigning weight $p_l/\sum_jp_j$ to direction $\boldsymbol u_l$. The corresponding normalized spatial correlation is
\begin{equation}\label{eq:phi}
\begin{aligned}
 \phi(\boldsymbol r)
 &=\sum_{l=1}^{L_p}\frac{p_l}{\sum_jp_j}
 e^{j\kappa\boldsymbol u_l^T\boldsymbol r}\\
 &=\int e^{j\kappa\boldsymbol u^T\boldsymbol r}\,\mu(d\boldsymbol u).
\end{aligned}
\end{equation}

The integral covers discrete paths and continuous angular spectra. Along the UE motion, $\phi_m(r)=\phi(r\mhat)$. A uniform azimuthal angular power spectrum gives $\phi_m(r)=J_0(\kappa r)$. For asymmetric spectra, including clustered delay line (CDL) ray distributions, the correlation can be complex, while its real part is even in $r$.
\subsection{Adjustment Target and Residual Tolerance}\label{sec:control-target}
The nominal compensating displacement is $D=v\tau$, estimated from UE motion and exchange timing. Exact restoration uses $q=D$. Otherwise we require $|D-q|\le\varepsilon$ with $0\le\varepsilon<\lambda/4$, motivated by the bound of Lemma~\ref{prop:cert}, which holds for any angular spectrum. The target must also be reachable before feedback, with switching, travel, and any mechanical reset fitting the schedule.
\subsection{Discrete Antenna Positioning}\label{sec:ports}
For discrete positioning, the UE provides $N$ candidate positions, each served by a calibrated element, with spacing $d$ between adjacent positions along the estimated motion direction. Element $0$ receives the forward signal, and element $j$ lies $jd$ behind it relative to the UE body. Switching to that element supplies adjustment $q=jd$. The same reference element, indexed by $0$, receives every forward transmission, so the aperture needs to span only the displacement $v\tau$ of one exchange, independently of the distance traveled during training. Minimizing the nominal residual gives
\begin{equation}\label{eq:select-control}
 j^\star=\underset{j\in\{0,\ldots,N-1\}}{\arg\min}|D-jd|,
 \quad r=D-j^\star d.
\end{equation}

Discrete positioning meets the residual tolerance when $|D-j^\star d|\le\varepsilon$.
The UE switches to $j^\star$ before $t_b$. Switching must fit the turnaround interval, and element responses must be calibrated. Exact restoration requires $D=j^\star d$, not merely $D\le(N-1)d$. A multi-antenna UE switches to a translated subarray in the same way. Coverage without gaps up to the aperture edge holds if $d\le2\varepsilon,\qquad D\le(N-1)d+\varepsilon$.
Otherwise the nearest-element residual is tested directly.
\subsection{Mechanical Counter-Movement and Reset}\label{sec:feasible}
A mechanical actuator moves the active antenna continuously opposite to the body motion. An isolated exchange with relative speed limit $V$, remaining travel $L$, and no overshoot has the displacement bound $0\le q\le\min\{V\tau,L,D\}$.
Exact cancellation in this isolated model requires $D\le L$ and $v\le V$. When either condition fails, projecting the target $D$ onto the admissible interval gives
\begin{equation}\label{eq:mechanical-control}
 q^\star=\min\{D,V\tau,L\},\qquad r=D-q^\star.
\end{equation}

For repeated exchanges, let $V$ and $a$ denote the relative speed and acceleration limits, $L$ be the available full rail stroke, $t_g$ be the guard time, $T$ the scheduled cycle period, and $t_p$ the available lead time before $t_f$.

For hold speed $u\in[0,v]$ relative to the UE body, start acceleration at $t_f-t_g-u/a$, hold $u$ over $[t_f-t_g,t_b+t_g]$, decelerate to rest, and reset to the starting rail position. Acceleration and deceleration thus occur outside $[t_f-t_g,t_b+t_g]$, so the antenna moves at the constant relative speed $u$ throughout the interval between the two transmissions, giving $q=u\tau$ and $r=(v-u)\tau$. The full stroke is
\begin{equation}\label{eq:stroke}
 S(u)=u(\tau+2t_g)+u^2/a.
\end{equation}

Including the rest-to-rest reset, the minimum cycle period is
\begin{equation}\label{eq:period}
 \begin{aligned}
 T_{\min}(u)&=\tau+2t_g+2u/a\\
 &\quad+\begin{cases}
 2\sqrt{S(u)/a},&S(u)\le V^2/a,\\
 S(u)/V+V/a,&S(u)>V^2/a.
 \end{cases}
 \end{aligned}
\end{equation}
The feasible hold-speed set for this protocol is
\begin{equation}\label{eq:mechanical-set}
 \begin{split}
 \mathcal U=\{u:\;&0\le u\le\min(v,V),\ S(u)\le L,\\
 &T_{\min}(u)\le T,\ t_g+u/a\le t_p\}.
 \end{split}
\end{equation}

Because $q$ counts only the displacement between $t_f$ and $t_b$, during which the antenna holds the speed $u$, the smallest hold speed meeting the residual budget is $u_{\rm req}=\max\{0,v-\varepsilon/\tau\}$. It is feasible precisely when $u_{\rm req}\in\mathcal U$, which then gives $q=u_{\rm req}\tau$ and $r=(v-u_{\rm req})\tau$. Increasing $u$ cannot repair a failed check because the resource requirements of this protocol increase with $u$. Exact compensation uses $u=v$, while nonbinding travel and timing reduce the residual test to $(v-V)_+\tau\le\varepsilon$. Reset returns the antenna to its initial position relative to the UE body, so rail travel does not accumulate across exchanges. Repeated operation must still account for actuator wear and motion power~\cite{wu2025globalma}, whereas discrete positioning only switches RF paths.
\subsection{Operating Decisions and Feasibility Boundaries}\label{sec:control-regions}
Algorithm~\ref{alg:position} tests the fixed, discrete, and mechanical options in this order, preferring discrete positioning when both adjusted modes are feasible.

For half-wavelength spacing and the limiting $\lambda/4$ budget, the three envelopes are
\begin{align}\label{eq:map-boundaries}
 v\tau&=\lambda/4,\nonumber\\
 v\tau&=(N-1)d+\lambda/4,\nonumber\\
 (v-V)_+\tau&=\lambda/4.
\end{align}

These are limiting envelopes. Operation uses $\varepsilon<\lambda/4$ and the full test~\eqref{eq:mechanical-set}. The mechanical envelope assumes nonbinding travel, lead-time, and reset constraints.

\begin{algorithm}[t]
\caption{Antenna adjustment for one BS-UE exchange}
\label{alg:position}
\small
\begin{algorithmic}[1]
\Require Estimated $v$, scheduled $\tau$, $\lambda,\varepsilon$, and UE hardware
\Require Discrete positioning: $N,d$ and switching time
\Require Mechanical: $V,a,L,t_g,t_p,T$
\Ensure Plan with nominal $q,r$, or an infeasibility report
\State Check $0\le\varepsilon<\lambda/4$. Set $D\gets v\tau$
\If{$D\le\varepsilon$}
 \State \Return Fixed-antenna plan with $q=0$, $r=D$
\EndIf
\If{discrete positioning is available and switching fits before $t_b$}
 \State $j^\star\gets\arg\min_{j\in\{0,\ldots,N-1\}}|D-jd|$
 \If{$|D-j^\star d|\le\varepsilon$}
  \State Use element $0$ for forward reception and element $j^\star$ for gradient feedback
  \State \Return Discrete plan with $q=j^\star d$, $r=D-q$
 \EndIf
\EndIf
\If{a mechanical actuator is available}
 \State $u\gets\max\{0,v-\varepsilon/\tau\}$
 \If{$u\le V$ and $S(u)\le L$}
  \If{$T_{\min}(u)\le T$ and $t_g+u/a\le t_p$}
   \State Plan acceleration from $t_f-t_g-u/a$
   \State Maintain speed $u$ through both transmissions and guard intervals
   \State Include deceleration and reset within the cycle
   \State \Return Mechanical plan with $q=u\tau$, $r=D-q$
  \EndIf
 \EndIf
\EndIf
\State \Return Infeasible for this tolerance and schedule
\end{algorithmic}
\end{algorithm}
\section{Theoretical Analysis}\label{sec:theory}
This section first establishes exact restoration and then quantifies how much of the descent direction survives a residual displacement. Weighted averages of path phases $e^{j\kappa\boldsymbol u^T\boldsymbol r}$ describe scalar correlations and alignment in fixed receive directions, and dual-basis path-mode precoding gives separate pathwise margins. Gaussian jitter attenuates each path contribution, and a systematic offset rotates its phase. General path-mode precoders and adaptive receive processing additionally require the matrix and conditional-alignment analyses below.

\subsection{Geometric Restoration and Implementation Feasibility}\label{sec:restore}
We first fix what exact restoration buys and what hardware it needs. Let $\mathcal I_f$ contain the forward variables, including $H_f$, $\delta$, and $P$.
\begin{proposition}[Exact channel and gradient restoration]\label{thm:restoration}
If the two transmissions share $H(\boldsymbol p)$, $\E[w\mid\mathcal I_f]=0$, and $\boldsymbol r_{\rm act}=\boldsymbol0$ (for instance, $q=v\tau$ without positioning error), then
\begin{equation}\label{eq:exact-restoration}
 H_b=H_f,\qquad \hat g-g=P(w),\qquad \E[\hat g\mid\mathcal I_f]=g .
\end{equation}
\end{proposition}
\begin{proof}
With $\boldsymbol r_{\rm act}=\boldsymbol0$, the backward position equals the forward position, $\boldsymbol p_b=\boldsymbol p_f$, so the shared spatial channel gives $H_b=H(\boldsymbol p_b)=H(\boldsymbol p_f)=H_f$. Substitution into~\eqref{eq:backward} and real linearity of $P$ give
\begin{equation}
 \hat g-g=P\bigl((H_b-H_f)^H\delta+w\bigr)=P(w),
\end{equation}
whose conditional mean vanishes.
\end{proof}
Restoring the antenna to its world position at the forward transmission removes channel aging without any statistical channel model or channel predictor~\cite{jiang2025csipred}. The backpropagated gradient estimate is then conditionally unbiased, so its conditional mean alignment with $g$ attains the ideal value $\norm{g}^2$.

Write $D=v\tau$ with $\tau>0$. The available modes must satisfy their switching, lead-time, and reset constraints.
\begin{corollary}[Hardware attainability of exact restoration]\label{cor:feasibility}
Under Proposition~\ref{thm:restoration}, exact restoration is attainable in each mode precisely when
\begin{equation}\label{eq:exact-hardware}
 \begin{aligned}
 D&\in\{0,d,\ldots,(N-1)d\} &&\text{(discrete)},\\
 V&\ge v,\quad L\ge D &&\text{(isolated motion)},\\
 v&\in\mathcal U,\quad u=v &&\text{(repeated motion)}.
 \end{aligned}
\end{equation}

Otherwise, nearest-element switching satisfies $|r|\le d/2$ when $D\le(N-1)d+d/2$, and the minimum isolated mechanical residual is $r_{\min}=\max\{0,(v-V)\tau,D-L\}$.
For repeated operation, $u_{\rm req}\in\mathcal U$ ensures $|r|\le\varepsilon$.
\end{corollary}

\begin{proof}
Discrete positioning attains $q=D$ exactly when $D=jd$ for an available index $j$. Isolated motion attains it when $\min\{D,V\tau,L\}=D$, i.e., $V\ge v$ and $L\ge D$. Repeated motion attains it when $u\tau=D$, i.e., $u=v$. Proposition~\ref{thm:restoration} then gives~\eqref{eq:exact-restoration}. Otherwise nearest-neighbor rounding gives the discrete bound, and
\begin{equation}
 D-\min\{D,V\tau,L\}=\max\{0,(v-V)\tau,D-L\}.
\end{equation}
Finally, $u_{\rm req}\in\mathcal U$ supplies the speed, stroke, lead time, and reset period with $(v-u_{\rm req})\tau\le\varepsilon$, and the reset restores the starting rail position, so the bound holds in every cycle.
\end{proof}
\subsection{Residual Correlation and Gradient Alignment}\label{sec:corr}
To quantify an imperfect adjustment, let $\F$ contain the input, label, current parameters, forward channel and noise, and $\delta$. Use the realized residual $\boldsymbol r_{\rm act}=\boldsymbol b_{\rm eff}+\boldsymbol e$.
\begin{assumption}[Scalar conditional mean]\label{ass:scalar}
The backward map $P$ is complex-linear, and
\begin{equation}
\begin{aligned}
 \E[H_b\mid\F,\boldsymbol r_{\rm act}]&=\phi(\boldsymbol r_{\rm act})H_f,\\
 \E[w\mid\F,\boldsymbol r_{\rm act}]&=0.
\end{aligned}
\end{equation}
\end{assumption}
Assumption~\ref{ass:scalar} is a modeling assumption, not a consequence of reciprocity: on average, the aged channel is the forward channel scaled by one complex factor, the spatial correlation. Its real part then controls mean-gradient alignment. This conditional-mean assumption is satisfied by the independent-innovation model in Assumption~\ref{ass:conv}(b). For finite ray clusters observed through an array, paths from different directions rotate by different phases, so the conditional mean of the aged channel is a matrix function of the forward channel, not a scalar multiple of it. The array analysis of Section~\ref{sec:array} does not use Assumption~\ref{ass:scalar}. Writing $\rho=\phi(\boldsymbol r_{\rm act})$ gives
\begin{align}\label{eq:mean}
 \E[\hat g\mid\F,\boldsymbol r_{\rm act}]&=\rho^*g,\nonumber\\
 \R\langle g,\E[\hat g\mid\F,\boldsymbol r_{\rm act}]\rangle
 &=\R\rho\,\norm{g}^2.
\end{align}

Thus, for $g\ne0$, a positive real part of the correlation makes the negative conditional mean of $\hat g$ a strict descent direction~\cite{twc}, as the strips in Fig.~\ref{fig:system} illustrate. The scalar and fixed-beam correlation bounds below follow from a common bound on the real part of a weighted path-phase average, which we state for a general probability measure.
\begin{lemma}[Weighted phase average]\label{prop:cert}
Let $\nu$ be a probability measure on unit vectors and $\hat{\boldsymbol t}$ a unit vector. Define $c_\nu=\sup_{\boldsymbol u\in\operatorname{supp}\nu}|\boldsymbol u^T\hat{\boldsymbol t}|$ and $s_\nu=(\int(\boldsymbol u^T\hat{\boldsymbol t})^2\nu(d\boldsymbol u))^{1/2}$, and write $A_\nu(r)=\R\int e^{j\kappa r\boldsymbol u^T\hat{\boldsymbol t}}\nu(d\boldsymbol u)$. Then
\begin{equation}\label{eq:correlation-lower-bounds}
\begin{aligned}
 A_\nu(r)&\ge 1-\tfrac12\kappa^2r^2s_\nu^2,
 &&r\ge0,\\
 A_\nu(r)&\ge\cos(\kappa c_\nu r),
 &&r\ge0,\quad\kappa c_\nu r\le\pi.
\end{aligned}
\end{equation}
Thus the real part of the average is strictly positive for $0\le r<\max\{\lambda/(4c_\nu),\lambda/(\sqrt2\pi s_\nu)\}$, with a zero denominator interpreted as an infinite radius. The $c_\nu$ radius is sharp: a single path along $\hat{\boldsymbol t}$ has $A_\nu(r)=\cos(\kappa r)$ and first zero $\lambda/4$.
\end{lemma}
\begin{proof}
The real part equals $\int\cos(\kappa r\,\boldsymbol u^T\hat{\boldsymbol t})\,\nu(d\boldsymbol u)$. Applying $\cos x\ge1-x^2/2$ under the integral gives the first bound. Since $|\boldsymbol u^T\hat{\boldsymbol t}|\le c_\nu$ on the support and cosine decreases on $[0,\pi]$, the integrand is at least $\cos(\kappa c_\nu r)$, which gives the second. Each bound is positive below its radius.
\end{proof}
Define $\sigma_m^2=\int(\boldsymbol u^T\mhat)^2\mu(d\boldsymbol u)\in[0,1]$, so $\sigma_m$ is the root-mean-square direction cosine along the motion. With $\nu=\mu$ and $\hat{\boldsymbol t}=\mhat$, $c_\nu\le1$ and $s_\nu=\sigma_m$. The first-lobe radius $r_0=\inf\{r>0:\R\phi_m(r)\le0\}$, with $\inf\varnothing=+\infty$, therefore satisfies
\begin{equation}\label{eq:cert}
 r_0\ \ge\ \max\bigl\{\lambda/4,\ \lambda/(\sqrt2\pi\sigma_m)\bigr\}.
\end{equation}
The second term is interpreted as $+\infty$ when $\sigma_m=0$.
Below this radius the real part of the correlation is positive for every angular spectrum. Beyond it, the correlation can become negative but need not, so the radius is a guarantee rather than a failure threshold.

Under Assumption~\ref{ass:scalar}, in the absence of offset and jitter, the nominal residual budget of Algorithm~\ref{alg:position} gives the mean-alignment margin $\R\rho\ge\cos(\kappa\varepsilon)>0$. With positioning errors, the corresponding averaged guarantee is given by Theorem~\ref{thm:jitter}. A known angular spectrum permits a deterministic residual budget below its own first zero $r_0$.
\subsection{Gaussian Jitter and Systematic Offset}\label{sec:jitter}
A fixed offset samples the spatial correlation at one displacement. Jitter averages it over the positioning-error distribution.
Write $\bar\rho=\E_{\boldsymbol e}[\phi(\boldsymbol b_{\rm eff}+\boldsymbol e)]$ for the mean spatial correlation.
\begin{theorem}[Jitter and offset]\label{thm:jitter}
Let $\mu$ be a normalized angular measure and $\boldsymbol e\sim\mathcal N(\boldsymbol0,\Sigma)$ be independent of the field and $\F$. Then
\begin{equation}\label{eq:gaussidentity}
 \bar\rho=\int e^{j\kappa\boldsymbol u^T\boldsymbol b_{\rm eff}}\,
 e^{-\frac12\kappa^2\boldsymbol u^T\Sigma\boldsymbol u}\,\mu(d\boldsymbol u),
\end{equation}
so jitter only scales the weight of each path and the offset only rotates its phase.
If $\boldsymbol b_{\rm eff}=\boldsymbol0$, then
\begin{equation}\label{eq:zero-offset-correlation}
 0<e^{-\kappa^2\lambda_{\max}(\Sigma)/2}\le\bar\rho\le1.
\end{equation}

If $\kappa\norm{\boldsymbol b_{\rm eff}}<\pi/2$, then
\begin{equation}\label{eq:biasbudget}
 \R\bar\rho\ \ge\ e^{-\kappa^2\lambda_{\max}(\Sigma)/2}\cos(\kappa\norm{\boldsymbol b_{\rm eff}})>0 .
\end{equation}

Under Assumption~\ref{ass:scalar},
\begin{equation}\label{eq:jitter-mean-gradient}
 \E[\hat g\mid\F]=\bar\rho^*g.
\end{equation}
\end{theorem}

\begin{proof}
Fubini's theorem and the Gaussian characteristic function, valid also for singular $\Sigma$,
\begin{equation}
 \E e^{j\kappa\boldsymbol u^T\boldsymbol e}=e^{-\kappa^2\boldsymbol u^T\Sigma\boldsymbol u/2}\in\bigl[e^{-\kappa^2\lambda_{\max}(\Sigma)/2},1\bigr],
\end{equation}
give~\eqref{eq:gaussidentity} and the zero-offset bounds. For $\kappa\norm{\boldsymbol b_{\rm eff}}<\pi/2$, $\cos(\kappa\boldsymbol u^T\boldsymbol b_{\rm eff})\ge\cos(\kappa\norm{\boldsymbol b_{\rm eff}})>0$ gives~\eqref{eq:biasbudget}, and averaging~\eqref{eq:mean} gives $\E[\hat g\mid\F]=\bar\rho^*g$.
\end{proof}

Zero mean alone does not preserve this sign. Other positioning-error distributions can yield negative correlation. A velocity-estimation bias $\varepsilon_b$ adds the persistent offset $\varepsilon_b\tau$, so the calibration budget for any angular spectrum becomes $|r|+|\varepsilon_b|\tau<\lambda/4$.
\subsection{Array and Beam-Resolved Alignment}\label{sec:array}
For an array, the backward product passes through receive processing, and the scalar model need not hold. For a fixed combiner, the array changes only how the paths are weighted, as the results below show. Let the receiver-side model first apply a linear combiner $M\in\mathbb C^{N_s\times N_r}$ and let $\boldsymbol d\in\mathbb C^{N_s}$ be the loss gradient at its output, so that $\delta=M^H\boldsymbol d$. The exact and received gradients with respect to the transmitted symbols are $H_f^HM^H\boldsymbol d$ and $H_b^HM^H\boldsymbol d$, and their inner product is $\boldsymbol d^HB_r\boldsymbol d$ with the aged operator $B_r=MH_fH_b^HM^H$.
At $r=0$, $B_0\succeq0$ for every channel. For $r\ne0$ the Hermitian part of a realized $B_r$ need not be positive semidefinite. We characterize its mean over the path gains. Consider the field-response model~\eqref{eq:multipath} with independent zero-mean circularly symmetric gains, $\E|\alpha_l|^2=p_l$, steering vectors with unit-modulus entries, and $M\boldsymbol a_{r,l}\ne\boldsymbol 0$ for some $l$ with $p_l>0$.

Define $\bar B_{\boldsymbol r}=\E[B_{\boldsymbol r}]$ and $\gamma_0=e^{-\kappa^2\lambda_{\max}(\Sigma)/2} \cos(\kappa\norm{\boldsymbol b_{\rm eff}})$.
\begin{theorem}[Array alignment]\label{prop:array}
Under the stated path-gain assumptions, let $M$ be fixed independently of the gains. If $\norm{\boldsymbol r}<\lambda/4$, then
\begin{equation}\label{eq:array-residual-margin}
 \Hm(\bar B_{\boldsymbol r})
 \succeq\cos(\kappa\norm{\boldsymbol r})\bar B_{\boldsymbol0}
 \succeq0.
\end{equation}

If $\boldsymbol e\sim\mathcal N(\boldsymbol0,\Sigma)$ is independent of the gains and $\kappa\norm{\boldsymbol b_{\rm eff}}<\pi/2$, then
\begin{equation}\label{eq:loewner}
 \Hm\bigl(\E_{\boldsymbol e}\bar B_{\boldsymbol b_{\rm eff}+\boldsymbol e}\bigr)\succeq\gamma_0\,\bar B_{\boldsymbol 0},
\end{equation}
\end{theorem}
Theorem~\ref{prop:array} bounds mean alignment in every fixed receive direction. The trace, by contrast, averages over directions and can be positive even when some directions have negative alignment.
\begin{proof}
The mean operator has the path expansion
\begin{equation}\label{eq:barB}
 \bar B_{\boldsymbol r}:=\E[B_{\boldsymbol r}]=N_t\sum_lp_l\,e^{-j\kappa\boldsymbol u_l^T\boldsymbol r}\,(M\boldsymbol a_{r,l})(M\boldsymbol a_{r,l})^H ,
\end{equation}

With $\boldsymbol h_l=M\boldsymbol a_{r,l}$, independent centered gains give $\E[\alpha_l\alpha_j^*]=p_l\mathbf 1_{\{l=j\}}$, which removes the cross terms between distinct paths and yields~\eqref{eq:barB}. At $\boldsymbol r=\boldsymbol0$,~\eqref{eq:array-residual-margin} holds with equality. For $\boldsymbol r\ne\boldsymbol0$, the Hermitian part replaces each phase by its cosine. Apply Lemma~\ref{prop:cert} to each path as a point mass, with $\hat{\boldsymbol t}=\boldsymbol r/\norm{\boldsymbol r}$ and scalar displacement $\norm{\boldsymbol r}$. Since $c_\nu\le1$ and $\kappa\norm{\boldsymbol r}<\pi/2$, this gives
\begin{equation}
 \Hm(\bar B_{\boldsymbol r})=N_t\sum_l p_l\cos(\kappa\boldsymbol u_l^T\boldsymbol r)\boldsymbol h_l\boldsymbol h_l^H\succeq\cos(\kappa\norm{\boldsymbol r})\bar B_{\boldsymbol0}.
\end{equation}
Under Gaussian error each coefficient is at least $e^{-\kappa^2\boldsymbol u_l^T\Sigma\boldsymbol u_l/2}\cos(\kappa\boldsymbol u_l^T\boldsymbol b_{\rm eff})\ge\gamma_0$, which proves~\eqref{eq:loewner}. 
\end{proof}
For a direction $\boldsymbol\psi$ in the combiner-output space, with effective receive beam $M^H\boldsymbol\psi$, define $P_{\boldsymbol\psi}=\sum_lp_l|\boldsymbol\psi^HM\boldsymbol a_{r,l}|^2$ and let $\mu_{\boldsymbol\psi}$ normalize these path weights. Write $c_{\boldsymbol\psi}$ and $s_{\boldsymbol\psi}$ for the maximum absolute and the root-mean-square projected direction cosines under this measure.
\begin{lemma}[Beam-resolved alignment]\label{lem:beam}
For $\norm{\boldsymbol\psi}=1$ and $P_{\boldsymbol\psi}>0$,
\begin{equation}\label{eq:beam}
 \boldsymbol\psi^H\Hm(\bar B_{\boldsymbol r})\boldsymbol\psi
 =N_tP_{\boldsymbol\psi}\R\phi_{\boldsymbol\psi}(\boldsymbol r).
\end{equation}

For $\boldsymbol r=r\mhat$, positivity holds when
\begin{equation}
 |r|<\max\left\{\frac{\lambda}{4c_{\boldsymbol\psi}},
 \frac{\lambda}{\sqrt2\pi s_{\boldsymbol\psi}}\right\},
\end{equation}
a zero denominator is interpreted as an infinite radius.
\end{lemma}
\begin{proof}
Taking the quadratic form of the Hermitian part in~\eqref{eq:barB} gives~\eqref{eq:beam}. Lemma~\ref{prop:cert} with $\nu=\mu_{\boldsymbol\psi}$ and $\hat{\boldsymbol t}=\mhat$ gives the radius. The Rayleigh quotient then gives $\lambda_{\max}(\Hm(\bar B_{\boldsymbol r}))\ge N_tP_{\boldsymbol\psi}\R\phi_{\boldsymbol\psi}(\boldsymbol r)$.
\end{proof}
Receive-combining directions dominated by paths nearly perpendicular to the motion therefore change slowly. Both results fix $M$ before averaging over the path gains. An MMSE receiver, however, computes its combiner from the observed forward channel, so its effective receive direction depends on the same channel realization that ages. Alignment must then be conditioned on the information used to form $\boldsymbol d$. Let $\mathcal X$ denote that information and suppose $B_{\boldsymbol r}$ and $B_{\boldsymbol0}$ are integrable.
\begin{proposition}[Conditional alignment criterion for adaptive receive processing]\label{prop:policy}
For $\nu\in\mathbb R$, the inequality
\begin{equation}\label{eq:adaptive-margin}
 \E\,\R(\boldsymbol d^HB_{\boldsymbol r}\boldsymbol d)
 \ge\nu\,\E(\boldsymbol d^HB_{\boldsymbol0}\boldsymbol d)
\end{equation}
holds for every bounded $\mathcal X$-measurable $\boldsymbol d$ if and only if
\begin{equation}\label{eq:adaptive-psd}
 \E[\Hm(B_{\boldsymbol r})-\nu B_{\boldsymbol0}\mid\mathcal X]\succeq0\quad\text{almost surely}.
\end{equation}
\end{proposition}
\begin{proof}
Conditioning on $\mathcal X$ writes the difference in~\eqref{eq:adaptive-margin} as $\E\bigl[\boldsymbol d^H\E[\Hm(B_{\boldsymbol r})-\nu B_{\boldsymbol0}\mid\mathcal X]\boldsymbol d\bigr]$, which is nonnegative under~\eqref{eq:adaptive-psd}. Otherwise, a countable dense set of unit vectors supplies a fixed $\boldsymbol v$ and an $\mathcal X$-measurable event $\mathcal E$ of positive probability on which the conditional quadratic form is negative, and $\boldsymbol d=\boldsymbol v\mathbf 1_{\mathcal E}$ violates~\eqref{eq:adaptive-margin}.
\end{proof}
Because the criterion must hold for every $\mathcal X$-measurable direction, it is conservative for a particular receiver, whose gradient need not occupy the unfavorable directions. Section~\ref{sec:paths} gives a realization-wise identity for adaptive processing under path-mode precoding, and Section~\ref{sec:eval} tests the MMSE UE.
\subsection{Nonconvex Convergence}\label{sec:conv}
Alignment now has to be turned into learning progress. Let $W_k\in\mathbb C^p$ be the transmitter parameters at update $k$ with the real inner product $\R\langle\cdot,\cdot\rangle$, and let $\hat g_k$ be the physically received minibatch gradient. All samples of a minibatch share the channel of that exchange.
\begin{assumption}[Smooth loss and conditional gradient moments]\label{ass:conv}
\textnormal{(a) Loss.} The objective $\Lc$, the loss averaged over data and channel realizations, is $\ell$-smooth and bounded below by $\Lc_{\inf}$, and the initial parameter $W_1$ is deterministic.

\textnormal{(b) Channel innovation.} The Gaussian positioning error is independent of $\F$. Conditional on $(\F,\boldsymbol e)$, write
\begin{equation}
\begin{aligned}
 H_b&=\phi H_f+\sqrt{1-|\phi|^2}\,\tilde H,\\
 \phi&=\phi(\boldsymbol b_{\rm eff}+\boldsymbol e),\qquad
 \boldsymbol e\sim\mathcal N(\boldsymbol0,\Sigma).
\end{aligned}
\end{equation}

The innovation $\tilde H$ is zero-mean, distributed as $H_f$, and independent of $(\F,\boldsymbol e,w)$. $P$ is complex-linear, and $\E[w\mid\F,\boldsymbol e]=0$.

\textnormal{(c) Gradient moments.} The ideal minibatch gradient is unbiased and obeys
\begin{equation}\label{eq:ideal-gradient-moments}
 \begin{aligned}
 \E[g_k\mid W_k]&=\nabla\Lc(W_k),\\
 \E[\norm{g_k-\nabla\Lc(W_k)}^2\mid W_k]&\le\sigma_s^2,\\
 \E[\norm{g_k}^2\mid W_k]&\le G^2.
 \end{aligned}
\end{equation}

The channel-innovation and backward-noise contributions satisfy
\begin{equation}\label{eq:backward-moments}
 \begin{aligned}
 \E[\norm{P(\tilde H^H\delta)}^2\mid\F]&\le G^2,\\
 \E[\norm{P(w)}^2\mid W_k]&\le\sigma_w^2.
 \end{aligned}
\end{equation}
\end{assumption}
These assumptions separate the mean learning direction from fluctuations caused by minibatch sampling, channel innovation, and backward noise. They form a matched statistical model and do not follow from reciprocity alone. The innovation model implies Assumption~\ref{ass:scalar}.
For residual $r$, offset $\boldsymbol b$, and jitter covariance $\Sigma$, define $\Delta=\Lc(W_1)-\Lc_{\inf}$ and
\begin{align}\label{eq:gamma}
 \gamma&=e^{-\kappa^2\lambda_{\max}(\Sigma)/2}\cos\bigl(\kappa(|r|+\norm{\boldsymbol b})\bigr),\nonumber\\
 M_\gamma&=\sigma_s^2+\sigma_w^2+(1-\gamma^2)G^2 .
\end{align}
\begin{theorem}[Position compensation and convergence]\label{thm:main}
Suppose Assumption~\ref{ass:conv} holds, $\ell>0$, $\Delta>0$, and $\kappa(|r|+\norm{\boldsymbol b})<\pi/2$. SGD with
\begin{equation}
 W_{k+1}=W_k-\eta\hat g_k,\qquad
 \eta=\min\Bigl\{\frac{\gamma}{\ell},\sqrt{\frac{2\Delta}{K\ell M_\gamma}}\Bigr\},
\end{equation}
satisfies
\begin{equation}\label{eq:main}
 \frac1K\sum_{k=1}^K\E\norm{\nabla\Lc(W_k)}^2\ \le\ \frac{2\ell\Delta}{K\gamma^2}+\frac{2}{\gamma}\sqrt{\frac{2\ell\Delta M_\gamma}{K}} .
\end{equation}

When $M_\gamma=0$, take $\eta=\gamma/\ell$.
\end{theorem}
For a target tolerance $\epsilon>0$ on the average expected squared gradient norm, the bound gives
\begin{equation}\label{eq:convergence-complexity}
 K=O\left(\frac{\ell\Delta}{\gamma^2}
 \max\left\{\frac1\epsilon,\frac{M_\gamma}{\epsilon^2}\right\}\right).
\end{equation}

The proof is given in Appendix~\ref{app:sgd}. Residual, offset, and jitter enter only through $\gamma$, so positioning error acts as an iteration tax. Without jitter or offset, a residual of $\lambda/8$ gives $\gamma=\cos(\pi/4)$ and doubles the leading factor $1/\gamma^2$ in~\eqref{eq:convergence-complexity}. The constant needs no knowledge of the angular spectrum, and with the spectrum $\gamma$ can be replaced by $\R\bar\rho$. The real part of $\bar\rho$ determines the expected decrease of the loss per update, while its imaginary part only enters the second-moment bound and thus limits the step size. The bound also covers joint training of both sides when Assumption~\ref{ass:conv}(a) and~(c) hold for the joint loss: the receiver differentiates its own layers exactly, so the stacked mean gradient estimate $(\bar\rho^*\nabla_W\Lc,\nabla_{W_{\rm rx}}\Lc)$ has drift at least $\R\bar\rho\norm{\nabla\Lc}^2$.
\subsection{Path-Mode Extension and Power Constraints}\label{sec:paths}
A resolved-path model provides a different route to analyzing channel-dependent receive processing. With a fixed path-mode precoder, the transmitter's last linear layer admits a gradient identity that holds for each channel realization even for adaptive UE processing.

Consider $L\le N_t$ resolvable paths with fixed directions and write
\begin{equation}\label{eq:path-channels}
 \begin{aligned}
 H_f&=U\operatorname{diag}(\beta_l)V^H,\\
 H_b&=U\operatorname{diag}(\beta_l e^{j\theta_l})V^H,\\
 \theta_l&=\kappa\boldsymbol u_l^T\boldsymbol r_{\rm act}.
 \end{aligned}
\end{equation}

Here $U\in\mathbb C^{N_r\times L}$ and $V\in\mathbb C^{N_t\times L}$ collect the receive and transmit path responses, $\beta_l$ includes the forward path gain and phase, and $V$ has full column rank. The last transmitter layer is $s=FCx,\qquad A=F^HV$,
where $x\in\mathbb C^m$ contains the input features, $C\in\mathbb C^{L\times m}$ is trainable, and $F\in\mathbb C^{N_t\times L}$ is a fixed precoder with nonsingular $A$. Gradients and norms of $C$ use the real Frobenius inner product and Frobenius norm.

Let $\xi$ denote backward noise in the gradient of $C$. Define $D_\theta=\operatorname{diag}(e^{-j\theta_l})$ and $\bar D=\operatorname{diag}(\zeta_l)$, where
\begin{equation}\label{eq:path-mean-phase}
 \begin{aligned}
 \zeta_l&=e^{-j\kappa\boldsymbol u_l^T\boldsymbol b_{\rm eff}}
 e^{-\kappa^2\boldsymbol u_l^T\Sigma\boldsymbol u_l/2}.
 \end{aligned}
\end{equation}
\begin{theorem}[Gradient feedback through resolved path modes]\label{thm:paths}
Under the path model and fixed precoder above:

\textnormal{(i)} For every realization, combiner $M$, and output gradient $\boldsymbol d$,
\begin{equation}\label{eq:path-gradient}
 \hat g=AD_\theta A^{-1}g+\xi,
\end{equation}
\textnormal{(ii)} Suppose Assumption~\ref{ass:conv}(a) and its unbiased minibatch and noise moment conditions in (c) hold for $C$. If $\boldsymbol e\sim\mathcal N(\boldsymbol0,\Sigma)$ is independent of the forward pass and $\E[\xi\mid\F,\boldsymbol e]=0$, then
\begin{equation}\label{eq:path-mean-gradient}
 \E[\hat g\mid C]=A\bar DA^{-1}\nabla\Lc(C).
\end{equation}
\textnormal{(iii)} In addition to (ii), suppose $F=V(V^HV)^{-1}$, $\kappa(|r|+\norm{\boldsymbol b})<\pi/2$, and $\norm{FCx}\le1$ along unprojected SGD iterates. Then
\begin{equation}\label{eq:path-alignment}
 \R\langle\nabla\Lc(C),\E[\hat g\mid C]\rangle
 \ge\gamma\norm{\nabla\Lc(C)}^2.
\end{equation}
and~\eqref{eq:main} holds with the same step-size rule and $M_\gamma=\sigma_s^2+\sigma_w^2$.
\end{theorem}
The dual-basis precoder removes mixing between path coordinates, turning the displacement error into separate pathwise phase rotations. It uses path directions rather than instantaneous gains, although poorly conditioned path responses can make the power requirement restrictive.
\begin{proof}
For the dual-basis precoder in (iii), $A=F^HV=(V^HV)^{-1}V^HV=I$.

For a fixed forward realization, set $\boldsymbol z=\operatorname{diag}(\beta_l^*)U^HM^H\boldsymbol d$.
Differentiating the last linear layer gives
\begin{equation}\label{eq:path-proof-gradient}
 \begin{aligned}
 g&=A\boldsymbol z x^H,\\
 \hat g-\xi&=AD_\theta\boldsymbol z x^H
 =AD_\theta A^{-1}g.
 \end{aligned}
\end{equation}

This proves (i) for fixed or channel-dependent receive processing, including MMSE, without averaging over $M$ or $\boldsymbol d$. Independence of the Gaussian positioning error gives $\E[D_\theta\mid\F]=\bar D$. Centered backward noise and the unbiased minibatch gradient then yield
\begin{equation}\label{eq:path-proof-mean}
 \E[\hat g\mid C]
 =A\bar DA^{-1}\E[g\mid C]
 =A\bar DA^{-1}\nabla\Lc(C),
\end{equation}
which proves (ii).

For (iii), $A=I$ and each diagonal coefficient satisfies
\begin{equation}\label{eq:path-proof-margin}
 \R\zeta_l
 =e^{-\kappa^2\boldsymbol u_l^T\Sigma\boldsymbol u_l/2}
 \cos(\kappa\boldsymbol u_l^T\boldsymbol b_{\rm eff})
 \ge\gamma.
\end{equation}

Applying this bound row by row to $\nabla\Lc(C)$ proves~\eqref{eq:path-alignment}. Since $D_\theta$ is unitary, the moment bounds give
\begin{equation}\label{eq:path-proof-moment}
 \E[\norm{\hat g}^2\mid C]
 \le\norm{\nabla\Lc(C)}^2+\sigma_s^2+\sigma_w^2.
\end{equation}

The smoothness descent argument in Appendix~\ref{app:sgd}, with this second-moment bound and alignment margin $\gamma$, gives the stated convergence bound.
\end{proof}
Projection onto a transmit-power constraint set changes the optimization problem. A channel-distorted gradient can satisfy the constrained stationarity condition even when the true gradient does not. The next result compensates for the known residual phase before projection.
Let $\mathcal S$ be a nonempty closed convex transmit-power constraint set, with normal cone $N_{\mathcal S}$. Define the phase correction and projected update by $\Theta=\operatorname{diag}(e^{j\kappa\boldsymbol u_l^T\boldsymbol r})$.
\begin{equation}\label{eq:power-projected-update}
 C_{k+1}=\Pi_{\mathcal S}(C_k-\eta\Theta\hat g_k).
\end{equation}

A fixed point of the deterministic mean update satisfies
\begin{equation}\label{eq:power-mean-fixed-point}
 C=\Pi_{\mathcal S}\bigl(C-\eta\Theta\bar D\nabla\Lc(C)\bigr)
\end{equation}
\begin{theorem}[Power-constrained path-mode training]\label{prop:power}
Assume Theorem~\ref{thm:paths}(ii), $F=V(V^HV)^{-1}$, and $\Sigma=\sigma^2I$.

\textnormal{(i)} If $\boldsymbol b=\boldsymbol0$ and the phase-corrected stochastic gradient oracle has uniformly bounded conditional second moment on $\mathcal S$, projected SGD with a suitable step size achieves an $O(K^{-1/2})$ bound on the average squared norm of the gradient of the constrained objective's Moreau envelope.

\textnormal{(ii)} If $\kappa\norm{\boldsymbol b}<\pi/2$, every fixed point~\eqref{eq:power-mean-fixed-point} satisfies
\begin{equation}\label{eq:power-stationarity-error}
 \begin{aligned}
 &\operatorname{dist}\bigl(0,\nabla\Lc(C)+N_{\mathcal S}(C)\bigr)\\
 &\quad\le\max_l|\sin(\kappa\boldsymbol u_l^T\boldsymbol b)|
 \norm{\nabla\Lc(C)}\\
 &\quad\le\sin(\kappa\norm{\boldsymbol b})\norm{\nabla\Lc(C)}.
 \end{aligned}
\end{equation}
\textnormal{(iii)} Without $\Theta$, (ii) holds with $\boldsymbol b$ replaced by $\boldsymbol r+\boldsymbol b$ whenever $\kappa\norm{\boldsymbol r+\boldsymbol b}<\pi/2$.
\end{theorem}
The proof is given in Appendix~\ref{app:power}. With isotropic Gaussian jitter and zero systematic offset, phase correction leaves a positive scalar multiple of the true gradient. Phase correction thus removes what the controller knows, the nominal residual. What remains is set by what it does not know, the offset $\boldsymbol b$, through the stationarity error in~\eqref{eq:power-stationarity-error}. The result applies to the last linear layer under path-mode precoding. Deeper layers and per-example normalization require their own Jacobians.

\begin{figure*}[htbp]\centering
\includegraphics[width=0.9\textwidth]{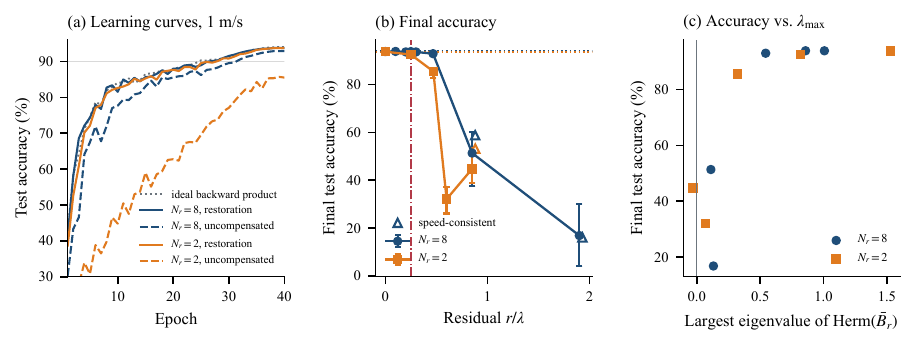}\par
\caption{CIFAR-10 with a ResNet-18 encoder over CDL-A (three seeds). (a) Learning curves at $1$~m/s (seed means). Solid: restoration. Dashed: no compensation ($r=0.47\lambda$). Dotted: ideal backward product. (b) Final accuracy versus residual. Error bars show the range over seeds, and dotted lines the accuracy with an ideal backward product. (c) Final accuracy against the largest eigenvalue of $\Hm(\bar B_r)$, a diagnostic rather than an accuracy predictor.}
\label{fig:cifar}
\end{figure*}

\subsection{Mechanical Cycle-Time Trade-off}\label{sec:cycle}
Theorem~\ref{thm:main} concerns progress per update. Training with mechanical compensation also incurs the cycle time in~\eqref{eq:period}. Near exact compensation, correlation loss is quadratic in residual displacement, whereas reducing the hold speed can shorten the cycle to first order. To examine this trade-off, assume time to target accuracy is proportional to the ratio of cycle time to the real part of the correlation
\begin{equation}\label{eq:partial-time-proxy}
 J(u)=\frac{T_{\min}(u)}{\R\phi_m((v-u)\tau)}.
\end{equation}
\begin{corollary}[Local benefit of partial compensation]\label{cor:partial}
Let $0<v\le V$ and suppose $T_{\min}(v)-T_{\min}(v-\varepsilon)\ge c\varepsilon$
for some $c>0$ and all sufficiently small $\varepsilon>0$. Then, for every angular spectrum, $J(v-\varepsilon)<J(v)$
for all sufficiently small $\varepsilon>0$.
\end{corollary}
Locally, the first-order cycle saving outweighs the second-order alignment loss, so a small residual improves this proxy. The conclusion does not depend on the exponent of the correlation in~\eqref{eq:partial-time-proxy}. With the $1/\gamma^2$ dependence of~\eqref{eq:convergence-complexity}, the alignment loss remains second order.
\begin{proof}
Lemma~\ref{prop:cert} gives $\R\phi_m(\epsilon\tau)\ge1-c'\epsilon^2$ with $c'=\kappa^2\sigma_m^2\tau^2/2$, so for small $\epsilon>0$
\begin{equation}
 J(v-\epsilon)\le\frac{T_{\min}(v)-c\epsilon}{1-c'\epsilon^2}<T_{\min}(v)=J(v).
\end{equation}
\end{proof}
When scheduling dominates the cycle, exact compensation adds no cycle-time penalty in this model. The timing advantage of partial compensation therefore requires mechanics to set the update period.

\section{Numerical Evaluation}\label{sec:eval}
The experiments test when restoration is needed, which residual errors disrupt learning, and how implementation time changes the comparison. CIFAR-10 evaluates classification performance, while synthetic binary classification tasks isolate propagation and positioning effects. All timing results use the stated implementation model.
\subsection{CIFAR-10 Setup}\label{sec:eval-cifar}
The main task is CIFAR-10, with a ResNet-18 encoder at the BS and a multilayer perceptron (MLP) classifier at the UE.

The channel follows the CDL-A profile of 3GPP TR 38.901~\cite{3gpp}, with $23$ clusters of $20$ rays, single polarization, and half-wavelength uniform linear arrays. The UE translates perpendicular to its array axis. Residuals are reported as $r/\lambda$. Physical implementation comparisons use $28$~GHz and a forward--backward interval of $5$~ms.

The encoder maps its $512$-dimensional feature to eight channel uses of four complex symbols, with unit average total transmit power per channel use, summed over the four transmit antennas. Unless stated otherwise, the link has eight receive and four transmit antennas, and the UE uses forward-channel MMSE combining followed by a $64\to256\to10$ MLP. The BS has no CSI. For the CSI-free UE, received-energy features $\log(1+|y|^2)$ feed a $64\to256\to256\to10$ MLP. A continuous trajectory advances $1.75\lambda$ per update, with the ray field redrawn every $100$ updates. The elapsed-time comparison instead advances each trajectory according to its actual exchange period. Within a block of $100$ updates, about $1.9$~m at $28$~GHz, the ray field is static. To isolate small-scale aging, consecutive blocks use independent fields at a fixed average signal-to-noise ratio. Large-scale path loss and blockage along the trajectory are not modeled. Within each forward or backward transmission, all channel uses for a batch of $128$ see the same flat-fading channel realization. Forward and backward noise standard deviations are $0.1$ and $0.03$, and feeding back the per-example gradient norm costs one real scalar per example.

Both sides use SGD with momentum, weight decay $5\times10^{-4}$, and a one-cycle learning-rate schedule over $40$ epochs with standard augmentation, peak step size $0.05$, and momentum $0.85$ to $0.95$. All configurations use the same three random seeds, and test images are evaluated after each epoch on a fixed set of independently drawn test channels. Ideal forward CSI is assumed for MMSE, without estimation overhead.
\subsection{Residual Displacement and Receiver Dependence}\label{sec:eval-res}
\begin{table}[!t]\centering\caption{Eigenspace intervention at $1$~m/s for the eight-antenna MMSE UE: final test accuracy, mean (range) over three seeds.}\label{tab:intervention}
\begin{tabular}{lc}\toprule
Backward signal & Accuracy (\%)\\\midrule
No compensation ($r=0.47\lambda$) & $92.95$ ($92.6$--$93.2$)\\
Positive eigenspace removed & $\mathbf{17.56}$ ($15.2$--$18.9$)\\
Random subspace removed, same dimension & $93.02$ ($92.8$--$93.2$)\\
Energy-matched rescaling & $92.95$ ($92.5$--$93.2$)\\
Energy-matched random removal & $92.59$ ($92.4$--$92.7$)\\
Restoration, positive eigenspace removed & $93.83$ ($93.7$--$94.0$)\\\bottomrule
\end{tabular}
\end{table} 

Fig.~\ref{fig:cifar}(a) and (b) show that exact restoration approaches the ideal-backward-product baseline for both receive arrays, whereas the effect of uncompensated motion depends strongly on the array. At walking speed, the eight-antenna MMSE UE retains high accuracy but needs $30.7$ epochs to reach $90\%$, compared with $26.7$ epochs with restoration, whereas the two-antenna UE never reaches $90\%$. With the per-update trajectory advance of Section~\ref{sec:eval-cifar} scaled to $1.82$ and $4.07$~m/s, uncompensated accuracy is $58.8\%$ and $16.1\%$, respectively. Discrete positioning stays within about $0.3$ percentage points of exact restoration at both speeds, although the $4.07$~m/s case exceeds the speed limit of the modeled actuator.

\begin{figure}[!t]\centering
\includegraphics[width=0.8\columnwidth]{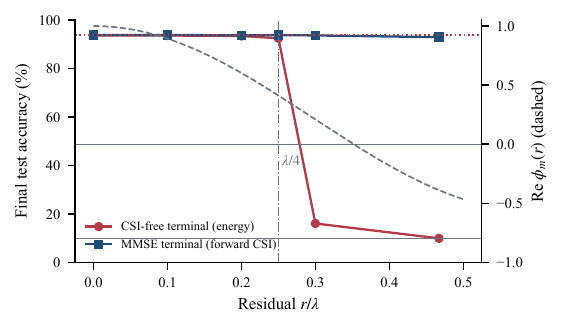}\par
\caption{CIFAR-10 residual sensitivity of the tested eight-antenna CSI-free and MMSE architectures. The $\lambda/4$ line marks a sufficient correlation bound valid for any angular spectrum, not a universal learning-failure threshold. Dotted: ideal backward product.}
\label{fig:csifree}
\end{figure}

The tested MMSE and CSI-free architectures have comparable accuracy with an ideal backward product but different residual tolerance (Fig.~\ref{fig:csifree}). The CSI-free UE remains near its baseline through $0.25\lambda$, then falls to $16.1\%$ at $0.3\lambda$. Uncompensated walking-speed training stays at chance. At $0.3\lambda$ the real part of the correlation is still about $0.21$, showing why a scalar correlation value alone does not predict this architecture's learning outcome.

\subsection{Why Some Arrays Tolerate Uncompensated Motion}
\begin{figure}[!t]\centering
\includegraphics[width=\columnwidth]{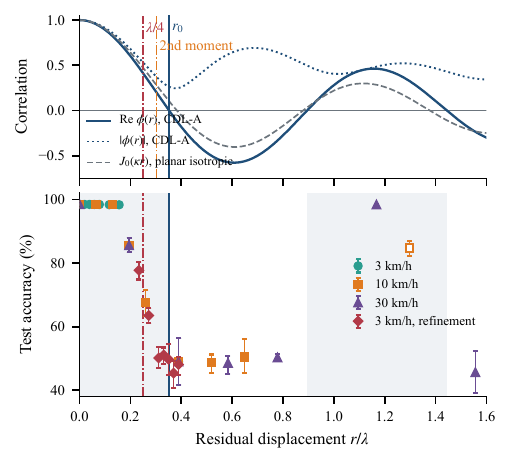}\par
\caption{Residual study on CDL-A (synthetic task). Top: real part of the correlation, $\R\phi_m(r)$, and the radii of Lemma~\ref{prop:cert}. Bottom: accuracy tracks the real part of the correlation, including recovery in the second positive lobe (shaded). Error bars: $95\%$ bootstrap intervals over eight seeds.}
\label{fig:residual}
\end{figure}

\begin{figure*}[t]\centering
\includegraphics[width=0.8\textwidth]{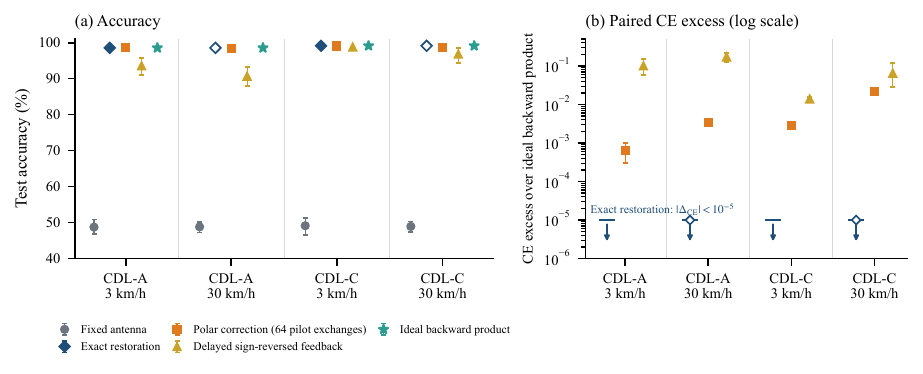}
\caption{Matched comparison on CDL-A/C at $3$ and $30$~km/h. Points are means over eight seeds with $95\%$ bootstrap intervals. Hollow markers: exact restoration at $30$~km/h, an idealized reference. (a) Test accuracy. (b) Paired cross-entropy (CE) excess over training with an ideal backward product. For exact restoration the whole interval lies within $\pm10^{-5}$ and is shown as an upper bound.}
\label{fig:benchsupp}
\end{figure*}

We estimate the mean combiner-output operator $\bar B_r$ from $300$ independent fields per geometry before training and normalize it by $\tr\bar B_0/N_t$. The largest eigenvalue of its Hermitian part tracks the two-antenna trend in Fig.~\ref{fig:cifar}(c), but weakly positive values coexist with poor eight-antenna accuracy. The failure of an uncompensated $2\times2$ link also shows that spatial stream overloading alone does not account for the small-array result.

For the intervention, the eigenvectors of $\Hm(\bar B_r)$ with positive eigenvalues are computed once from this $300$-field estimate at $r=0.47\lambda$, and before every backward transmission each example's combiner-output gradient $\boldsymbol d$ is replaced by its component orthogonal to them. The controls remove a random subspace of equal dimension, or rescale the unmodified or randomly projected signal to the norm left by the targeted removal. The restoration control applies the same projection at $r=0$. At walking speed, the targeted removal reduces uncompensated accuracy from $92.95\%$ to $17.6\%$, whereas removing a random subspace of equal dimension, the two energy-matched controls, and the same removal under restoration all stay near $93\%$ (Table~\ref{tab:intervention}).
These controls support a directional explanation beyond signal-energy loss.

\subsection{Channel Profiles and Mobility Speeds}

A synthetic binary classification task probes propagation-profile and mobility effects using CDL-A and CDL-C ($24$ clusters) at $3$ and $30$~km/h. Each example uses an independent field, with shared ray phases for its forward and backward transmissions.

A complex linear encoder maps three independent $\pm1$ inputs to two transmit streams, with $\norm{W}_F\le1/\sqrt3$. A four-antenna CSI-free receiver applies $\tanh(|y_i|^2)$ and logistic regression to the nonlinear label $\mathbf 1[x_0x_1>0]$. Forward and backward noise standard deviations are $0.1$ and $0.03$. Adam uses step sizes of $0.025$ and $0.06$ for the encoder and the receiver, respectively, batch size $32$, and $120$ updates. Evaluation uses $8192$ examples. All methods share eight seeds, eight frequency tones per seed, data, initialization, and noise, with per-example norm feedback for over-the-air gradient feedback. The encoder weight matrix is projected onto its Frobenius-norm ball after each update. Receiver gradients are computed locally.

\emph{Accuracy tracks the real part of the correlation.} The CDL-A residual sweep uses constant backward scaling, speeds of $3$, $10$, and $30$~km/h, and compensation ratios $q/(v\tau)$ from $0.5$ to $1$, together with fixed antennas. Across the $24$ mobility configurations, the precomputed $\R\phi_m(r)$ has a Spearman rank correlation of $0.953$ with accuracy, although accuracy is nonmonotonic in the residual (Fig.~\ref{fig:residual}). A residual displacement of $1.17\lambda$ in the second positive lobe reaches $98.35\%$, whereas a negative-lobe case is near chance and a first-lobe point with similar correlation magnitude reaches only $67.7\%$. Magnitude alone is insufficient.

\begin{figure}[t]\centering
\includegraphics[width=0.8\columnwidth]{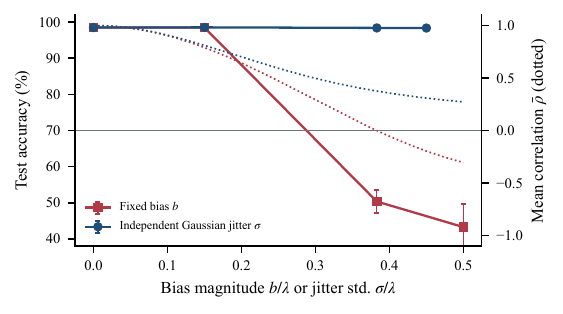}
\caption{Fixed bias and independent Gaussian rail jitter in the planar isotropic field (synthetic task). Dotted curves show their mean correlations on the right axis. The horizontal scale is bias magnitude or jitter standard deviation. Error bars: $95\%$ bootstrap intervals over eight seeds.}
\label{fig:bias}
\end{figure}
\emph{Comparison with digital alternatives.} We compare an uncompensated fixed antenna, exact restoration, pilot-based polar correction~\cite{twc}, delayed sign-reversed gradient feedback at a later negative correlation extremum~\cite{wcl}, and an ideal backward product without backward noise. Polar correction estimates the aging operator from $64$ bidirectional pilot exchanges, with two pilot probes per direction on each of eight tones, and applies its unitary polar factor.

Fig.~\ref{fig:benchsupp} places restoration at the ideal-backward-product baseline across the tested channel profiles and speeds, while the fixed antenna remains near chance. Polar correction approaches this performance but uses $2048$ additional pilot tone uses per run. Delayed feedback adds $1.1$ to $3.2$~ms of feedback delay per exchange and loses accuracy. At $30$~km/h, exact restoration requires $4.2$~cm of adjustment in $5$~ms and serves only as an idealized reference.
\subsection{Systematic Offset Versus Random Jitter}
The synthetic task next separates systematic offset from random positioning error in a planar isotropic field with a four-element array. After compensating an initial displacement of $0.5\lambda$, we add either fixed bias or independent Gaussian rail jitter with matched root-mean-square displacement. Training retains the compact-task settings but uses constant backward scaling $1/16$ and inverse scaling after reception, rather than per-example norm feedback.

Fig.~\ref{fig:bias} separates errors with the same root-mean-square displacement: at $0.383\lambda$, fixed bias gives $50.3\%$ accuracy, while Gaussian jitter retains $98.42\%$. The contrast is consistent with the mean-correlation mechanism in Theorem~\ref{thm:jitter} and supports separate calibration and jitter budgets.
\subsection{Feasible Adjustment and Elapsed Training Time}\label{sec:eval-hardware}
At $28$~GHz, the $30$~km/h case requires nine half-wavelength-spaced positions, within the $16$-position aperture considered here (Fig.~\ref{fig:system}, markers). At $1$~m/s with $a=400$~m/s$^2$, exact mechanical compensation needs a $7.5$~mm stroke and an $18.7$~ms cycle. Allowing a residual of $\lambda/8$ reduces them to $5.0$~mm and $15.7$~ms. Discrete positioning also requires the receiving and transmitting elements to be phase-calibrated. In the synthetic task, a common RF phase-calibration error of $30^\circ$ leaves accuracy unchanged, whereas $45^\circ$ reduces test accuracy by $3$ to $6$ percentage points.

\begin{figure*}[htbp]\centering
\includegraphics[width=\textwidth]{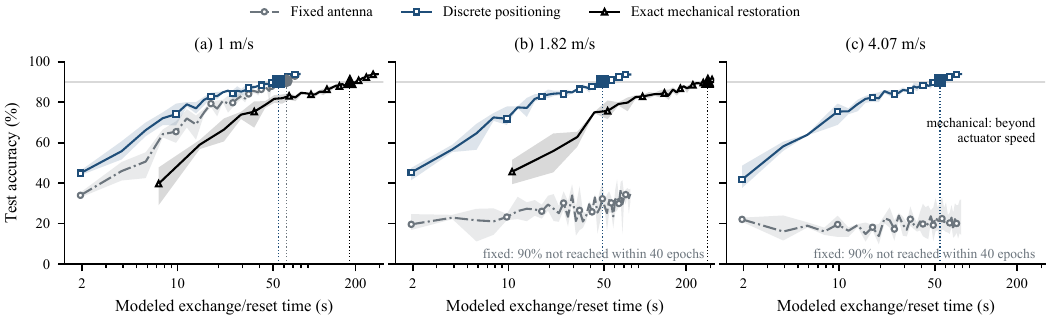}
\caption{CIFAR-10 accuracy versus modeled exchange-and-reset time, with trajectories regenerated at each implementation's exchange period. The eight-antenna MMSE UE is used. Curves show the mean and range over three seeds, and the gray line marks $90\%$. Filled markers: first epoch at which the seed-mean accuracy reaches $90\%$ (Table~\ref{tab:budget}).}
\label{fig:budget}
\end{figure*}

For the timing comparison, CIFAR-10 trajectories are regenerated at each implementation's exchange period, using the eight-antenna MMSE UE and the same three seeds. Fixed antennas and discrete positioning use a $5$~ms period. Exact mechanical restoration uses~\eqref{eq:period} with $V=2.5$~m/s and is excluded above that actuator speed limit.

Discrete positioning reaches $90\%$ first at every tested speed under the exchange-and-reset model (Table~\ref{tab:budget}, Fig.~\ref{fig:budget}). Exact mechanical restoration improves gradient fidelity, but its reset time lengthens the modeled time to target accuracy. The fixed antenna misses $90\%$ within $40$ epochs at higher speeds. The reported time includes forward--backward exchanges and antenna reset, not measured end-to-end latency. Computation, pilot, and feedback overheads are not separately added.

\begin{table}[!t]\centering\caption{Modeled exchange-and-reset time to $90\%$ test accuracy for an eight-antenna MMSE UE. The target is the first epoch at which seed-mean accuracy reaches $90\%$.}\label{tab:budget}
\begin{tabular}{lccc}\toprule
 & $1$~m/s & $1.82$~m/s & $4.07$~m/s\\\midrule
Fixed antenna & $62.4$~s & not reached$^{a}$ & not reached$^{a}$\\
Discrete pos. & $54.6$~s & $48.8$~s & $54.6$~s\\
Exact mechanical & $181.9$~s & $287.5$~s & exceeds limit$^{b}$\\\bottomrule
\end{tabular}\\[2pt]
\footnotesize $^{a}$Within $40$ epochs. Final accuracy $34.7\%$ and $20.5\%$. $^{b}$The speed exceeds $V$.
\end{table}
\section{Conclusion and Future Directions}
Movable-antenna control can compensate passive UE movement in over-the-air SL. For a stationary BS and a field quasi-static within each exchange, it restores the forward channel and thereby the backward product required for backpropagation. Under imperfect restoration, the tolerable residual depends on systematic offset, Gaussian jitter, and receive processing, and our scalar, array, and path-mode analyses give the corresponding alignment and learning guarantees. Exact restoration is not always the best choice for time: a small residual shortens the mechanical cycle to first order at only a second-order loss in correlation, and discrete positioning reaches the target accuracy first in the tested timing model. Compensation should therefore be designed jointly with receiver processing, positioning accuracy, and the update schedule.

Moving scatterers, rotation, mutual coupling~\cite{xu2026macoupling}, and multiuser scheduling remain untested. Prototype tests should measure motion-estimation error, switching calibration, actuator wear, and full computation and communication latency. These measurements would establish the benefit of repeatable compensation beyond the modeled exchange-and-reset time.
\appendices
\section{Proof of Theorem~\ref{thm:main}}\label{app:sgd}
\begin{lemma}[Mean and variance of the gradient estimate under channel aging]\label{lem:var}
Under Assumption~\ref{ass:conv}, $\E[\hat g_k\mid W_k]=\bar\rho^*\nabla\Lc(W_k)$ with $\bar\rho$ of~\eqref{eq:gaussidentity}, and
\begin{equation}\label{eq:var}
 \E\bigl[\norm{\hat g_k-\bar\rho^*\nabla\Lc(W_k)}^2\mid W_k\bigr]\le\sigma_s^2+\sigma_w^2+(1-|\bar\rho|^2)G^2.
\end{equation}
\end{lemma}
\begin{proof}
With $v_k=\nabla\Lc(W_k)$, Assumption~\ref{ass:conv}(b) gives
\begin{equation*}
 \hat g_k=\phi^*g_k+\sqrt{1-|\phi|^2}\,P(\tilde H^H\delta)+P(w),
\end{equation*}
whose last two terms have zero conditional mean. Independence of the positioning error and the forward pass gives $\E[\hat g_k\mid W_k]=\bar\rho^*v_k$. In the deviation $\bar\rho^*(g_k-v_k)+(\phi-\bar\rho)^*g_k+\sqrt{1-|\phi|^2}P(\tilde H^H\delta)+P(w)$ all cross terms vanish, so the moment bounds give
\begin{equation*}
 \E[\norm{\hat g_k-\bar\rho^*v_k}^2\mid W_k]\le|\bar\rho|^2\sigma_s^2+\sigma_w^2+(1-|\bar\rho|^2)G^2,
\end{equation*}
which implies~\eqref{eq:var} because $|\bar\rho|\le1$.
\end{proof}
\begin{proof}[Proof of Theorem~\ref{thm:main}]
Theorem~\ref{thm:jitter} and the offset triangle inequality give $\R\bar\rho\ge\gamma>0$ and $1-|\bar\rho|^2\le1-\gamma^2$, so Lemma~\ref{lem:var} bounds the variance by $M_\gamma$. With $v_k=\nabla\Lc(W_k)$, the drift is $\R\bar\rho\norm{v_k}^2$ and $\E[\norm{\hat g_k}^2\mid W_k]\le|\bar\rho|^2\norm{v_k}^2+M_\gamma$, so smoothness and $\eta\le\gamma/\ell\le\R\bar\rho/(\ell|\bar\rho|^2)$ give
\begin{equation*}
 \E[\Lc(W_{k+1})\mid W_k]\le\Lc(W_k)-\frac{\eta\gamma}{2}\norm{v_k}^2+\frac{\ell\eta^2}{2}M_\gamma .
\end{equation*}
Summing over $k$ and using the lower bound on the objective,
\begin{equation*}
 \frac1K\sum_{k=1}^K\E\norm{v_k}^2\le\frac{2\Delta}{K\eta\gamma}+\frac{\ell\eta M_\gamma}{\gamma}.
\end{equation*}
With $\eta=\min\{\gamma/\ell,\sqrt{2\Delta/(K\ell M_\gamma)}\}$, the first term is at most $2\ell\Delta/(K\gamma^2)+\sqrt{2\ell\Delta M_\gamma/K}/\gamma$ and the second at most $\sqrt{2\ell\Delta M_\gamma/K}/\gamma$, which gives~\eqref{eq:main}. This is the argument of~\cite[Th.~4.8]{bottou2018} with geometric drift and variance bounds.
\end{proof}
\section{Proof of Theorem~\ref{prop:power}}\label{app:power}
\begin{proof}
For zero systematic offset, phase correction gives
\begin{equation}\label{eq:power-unbiased-mean}
 \E[\Theta\hat g_k\mid C_k]
 =e^{-\kappa^2\sigma^2/2}\nabla\Lc(C_k).
\end{equation}
Thus the corrected oracle is unbiased for a positive multiple of the objective gradient.

Let $c=e^{-\kappa^2\sigma^2/2}$. Under isotropic jitter every path coefficient has magnitude $c$, so the known residual-phase correction gives $\Theta\bar D=c\operatorname{diag}(e^{-j\kappa\boldsymbol u_l^T\boldsymbol b})$, and zero offset makes the corrected oracle unbiased for $c\nabla\Lc$.

For part (i), write $f=\Lc$, $v_k=\Theta\hat g_k/c$ with $\E[\norm{v_k}^2\mid C_k]\le B^2$, $F=f+I_{\mathcal S}$ with $I_{\mathcal S}$ the indicator function of $\mathcal S$ (zero on $\mathcal S$ and $+\infty$ elsewhere), and $\lambda_{\rm M}=1/(2\ell)$ for the Moreau envelope $F_{\lambda_{\rm M}}$. With $y=\operatorname{prox}_{\lambda_{\rm M} F}(C)$, smoothness and proximal optimality give $\langle\nabla f(C),C-y\rangle\ge\norm{C-y}^2/(4\lambda_{\rm M})$, and nonexpansiveness of the projection then gives, for $C^+=\Pi_{\mathcal S}(C-\alpha v)$,
\begin{equation*}
 \E[F_{\lambda_{\rm M}}(C^+)\mid C]\le F_{\lambda_{\rm M}}(C)-\frac{\alpha}{4}\norm{\nabla F_{\lambda_{\rm M}}(C)}^2+\frac{\alpha^2B^2}{2\lambda_{\rm M}}.
\end{equation*}
Summing over $K$ iterations with $\alpha=\eta c\propto K^{-1/2}$ proves part (i).

For part (ii), put $\theta_l=\kappa\boldsymbol u_l^T\boldsymbol b$ and $A=c\operatorname{diag}(e^{-j\theta_l})$. At a fixed point, $-A g\in N_{\mathcal S}(C)$, where $g=\nabla f(C)$. Let $m=\min_l\cos\theta_l>0$. Since the normal cone is closed under nonnegative scaling,
\begin{equation*}
\begin{aligned}
 \operatorname{dist}(0,g+N_{\mathcal S}(C))&\le\norm{g-(m/c)Ag}\\
 &\le\sqrt{1-m^2}\,\norm{g}=\max_l|\sin\theta_l|\,\norm{g}.
\end{aligned}
\end{equation*}
The angle bound gives the final inequality in the theorem. Equality is attained on a complex unit disk with $f(z)=\R z$ and fixed point $z=-e^{-j\theta}$, for which the normal-cone distance is $|\sin\theta|$. Omitting $\Theta$ adds the known residual phase, proving part (iii).
\end{proof}
\input{appendix_ext}

\bibliographystyle{IEEEtran}
\bibliography{refs}
\end{document}

%% file: appendix_ext.tex
\section{Acceleration-Limited Reachability and the Mechanical Cycle}\label{app:accel}
Section~\ref{sec:feasible} uses a hold-speed protocol. This appendix states the largest displacement an acceleration-limited actuator can realize within one exchange when it starts from rest, and shows the resulting cycle period and stroke.

Let $u(t)$ be the antenna velocity relative to the UE body, positive opposite to the motion, with $|u|\le V$ and $|\dot u|\le a$, starting from rest at $t_f-t_p$. The compensation over the exchange is $q=\int_{t_f}^{t_b}u(t)\,dt$.
\begin{proposition}[Acceleration-limited reachability]\label{prop:accel}
Let $u_0=\min\{V,a t_p\}$. The largest reachable compensation is
\begin{equation}\label{eq:qmax}
 q_{\max}=\begin{cases}u_0\tau+\tfrac12a\tau^2,& V-u_0\ge a\tau,\\[2pt] V\tau-\dfrac{(V-u_0)^2}{2a},&\text{otherwise.}\end{cases}
\end{equation}
If the travel satisfies $L\ge u_0^2/(2a)+q_{\max}+\min\{V,u_0+a\tau\}^2/(2a)$, which accommodates the lead and stopping distances, every $q\in[0,q_{\max}]$ is reachable and the smallest residual is $(v\tau-q_{\max})_+$.
\end{proposition}
\begin{proof}
From rest at $t_f-t_p$, the bounds give $u(t_f+s)\le\min\{V,u_0+as\}$ for $s\in[0,\tau]$, and accelerating at $a$ until $V$ attains this envelope. Integrating it over $[0,\tau]$ gives~\eqref{eq:qmax}. Scaling a feasible velocity profile by $c\in[0,1]$ preserves feasibility and scales $q$ by $c$.
\end{proof}
With full lead time, $t_p\ge V/a$, the bound reduces to $q_{\max}=V\tau$, the isolated-exchange limit of Section~\ref{sec:feasible}.

\begin{figure*}[t]\centering
\includegraphics[width=\textwidth]{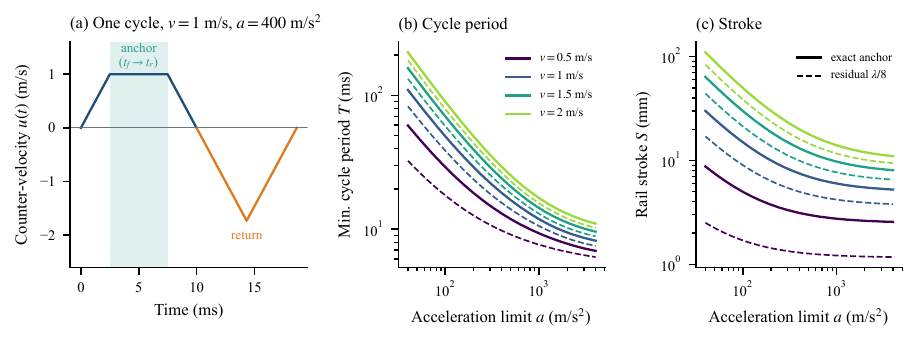}
\caption{Repeated mechanical compensation at $28$~GHz ($\lambda=1.07$~cm), $\tau=5$~ms, and $V=2.5$~m/s. (a) Antenna velocity over one cycle. (b) Minimum cycle period~\eqref{eq:period} and (c) rail stroke~\eqref{eq:stroke} versus the acceleration limit, for exact compensation (solid) and a residual of $\lambda/8$ (dashed).}
\label{fig:protocol}
\end{figure*}

Fig.~\ref{fig:protocol} illustrates the repeated protocol at $28$~GHz with $\tau=5$~ms and $V=2.5$~m/s. Exact compensation of a $1$~m/s UE with $a=400$~m/s$^2$ needs a $7.5$~mm stroke and an $18.7$~ms cycle. Allowing a residual of $\lambda/8$ lowers the hold speed to $0.73$~m/s, the stroke to $5.0$~mm, and the cycle to $15.7$~ms. At $2$~m/s the exact stroke grows to $20$~mm and the cycle to $29.2$~ms, so the cycle, not the exchange, sets the update rate.

\section{Operating Regions}\label{app:regions}
\begin{figure}[t]\centering
\includegraphics[width=\columnwidth]{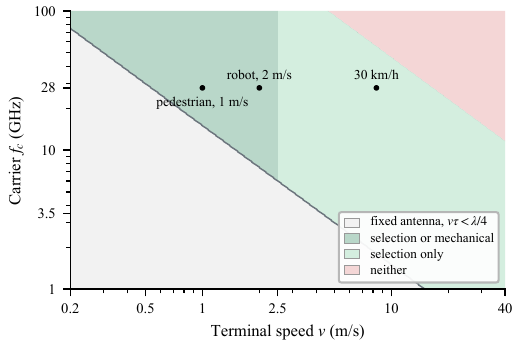}
\caption{Limiting operating regions for $\tau=5$~ms, $16$ candidate positions at half-wavelength spacing, and $V=2.5$~m/s.}
\label{fig:regions}
\end{figure}

Fig.~\ref{fig:regions} draws the limiting envelopes~\eqref{eq:map-boundaries} in the plane of UE speed and carrier frequency for $\tau=5$~ms, $N=16$ candidate positions at half-wavelength spacing, and an actuator speed limit $V=2.5$~m/s. Below the lower line the fixed antenna already keeps $v\tau<\lambda/4$. In the region marked ``discrete or mechanical'' both implementations meet the limiting budget, and in the region marked ``discrete positioning only'' the required speed exceeds $V$ while the displacement still lies within the candidate aperture. At these parameters no operating point is reachable by mechanical compensation alone. Mechanical compensation remains the choice when exact restoration is required, since discrete positioning leaves $|r|\le d/2$, and when the UE provides travel but no element aperture. The markers place a pedestrian at $1$~m/s and a robot at $2$~m/s at $28$~GHz, and a UE at $30$~km/h. The envelopes assume nonbinding travel, lead time, and reset period. Actual operation uses $\varepsilon<\lambda/4$ and the full test~\eqref{eq:mechanical-set}.

\section{Convergence Rate Versus the Real Part of the Correlation}\label{app:conv}
The mean drift in~\eqref{eq:mean} scales with the real part of the correlation, which suggests that, near a fixed step size, the number of updates to a target scales inversely with it. The bound of Theorem~\ref{thm:main} carries the more conservative leading factor $1/\gamma^2$. We trained the synthetic task of Section~\ref{sec:eval} on CDL-A with residuals $r/\lambda\in\{0,0.05,\dots,0.30\}$, for which $\R\phi_m$ ranges from $1$ to $0.21$, for $600$ updates with eight seeds and recorded the test accuracy after every update. The encoder used plain SGD with a fixed step $\eta=0.005$, the largest step for which the median time to $90\%$ accuracy at $r=0$ was at least $40$ updates on four separate calibration seeds. A second configuration used Adam. We regress the logarithm of the median number of updates to $90\%$ accuracy, $K_{90}$, on $\log\R\phi_m$.

\begin{figure*}[t]\centering
\includegraphics[width=\textwidth]{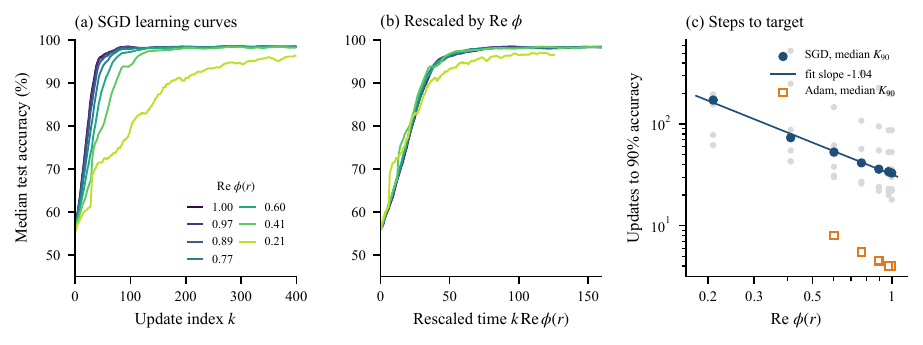}
\caption{Convergence study on CDL-A (synthetic task, eight seeds). (a) SGD learning curves. (b) The same curves against the rescaled time $k\,\R\phi_m$. (c) Updates to $90\%$ accuracy scale as $(\R\phi_m)^{-1.04}$ for SGD.}
\label{fig:conv}
\end{figure*}

This study tests the time-rescaling prediction rather than the bound itself, whose assumptions the per-example channel realizations, the encoder norm projection, and the Adam-trained receiver do not meet. The SGD slope is $-1.04$ ($R^2=0.993$, bootstrap interval over seeds $[-1.23,-0.68]$), and the learning curves collapse when plotted against $k\,\R\phi_m$ (Fig.~\ref{fig:conv}(b)). Median $K_{90}$ grows from $33$ updates at $r=0$ to $73.5$ at $\R\phi_m=0.41$ and $172.5$ at $\R\phi_m=0.21$. Using $|\phi_m|$ instead of $\R\phi_m$ fits worse ($R^2=0.970$, slope $-1.62$), consistent with the role of the real part in the mean drift. Adam is about eight times faster at small residuals but degrades as the correlation weakens. After $600$ updates it reaches $85.8\%$, $73.3\%$, and $52.7\%$ at $\R\phi_m=0.60$, $0.41$, and $0.21$, against $97.4\%$, $98.1\%$, and $87.1\%$ for SGD. Scaling the Adam step of the encoder by $\R\phi_m$ raises its final accuracy to $95.3\%$, $80.9\%$, and $66.5\%$ at $\R\phi_m=0.60$, $0.41$, and $0.21$ but does not close the gap to SGD. Table~\ref{tab:conv} lists the medians behind Fig.~\ref{fig:conv}(c). Seeds that do not reach $90\%$ within $600$ updates enter the median as censored values above the budget. The finite-budget accuracy losses inside the first lobe in Section~\ref{sec:eval} therefore come from the adaptive optimizer rather than from a lost mean descent direction.

\begin{table}[t]\centering\caption{Convergence study on CDL-A: median updates to $90\%$ accuracy over eight seeds, with the number of seeds that do not reach $90\%$ within $600$ updates (SGD).}\label{tab:conv}
\begin{tabular}{cccccc}\toprule
$r/\lambda$ & $\R\phi_m$ & SGD $K_{90}$ & censored & Adam $K_{90}$\\\midrule
$0$ & $1.000$ & $33$ & $0$ & $4$\\
$0.05$ & $0.973$ & $34$ & $0$ & $4$\\
$0.10$ & $0.893$ & $36$ & $0$ & $4.5$\\
$0.15$ & $0.767$ & $41.5$ & $0$ & $5.5$\\
$0.20$ & $0.603$ & $53$ & $1$ & $8$\\
$0.25$ & $0.412$ & $73.5$ & $0$ & $>600$\\
$0.30$ & $0.209$ & $172.5$ & $2$ & $>600$\\\bottomrule
\end{tabular}\end{table}

\section{Split Network on a Digit Dataset}\label{app:digits}
This study trains a multi-layer split network on the ten-class $8\times8$ handwritten-digit dataset ($1797$ images, $1200$ for training and $597$ for testing per seed). The BS-side model ($64\to64$ tanh $\to$ four complex symbols with unit power) transmits over an $8\times4$ CDL-A channel to a walking UE at $1$~m/s that advances $1.75\lambda$ per exchange. The UE applies MMSE combining with ideal knowledge of the forward channel, followed by an $8\to64$ tanh $\to10$ classifier with cross-entropy loss. The BS has no CSI and receives its gradient only through the reciprocal backward transmission with per-example norm feedback. The UE moves along a continuous trajectory through a ray field that is redrawn every $100$ updates, and each minibatch of $32$ shares one channel. Both sides use SGD with step size $0.05$ for $1500$ updates, and eight seeds pair all configurations. The network has real parameters, so its local backward map is only real-linear, and the channel-dependent combiner lies outside the fixed-combiner setting of Theorem~\ref{prop:array}. The study therefore tests the qualitative predictions outside the scope of the theorems.

\begin{figure*}[t]\centering
\includegraphics[width=\textwidth]{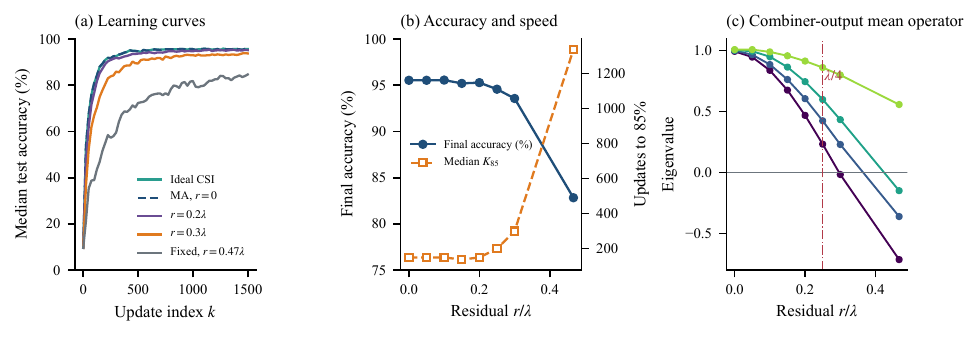}
\caption{Split network on the digit dataset over an $8\times4$ CDL-A channel with continuous UE trajectories. (a) Median test accuracy for an ideal backward product, exact restoration, two residuals, and the fixed antenna ($r=v\tau=0.47\lambda$). (b) Final accuracy and median updates to $85\%$ versus residual (evaluated every $25$ updates, and one of eight fixed-antenna seeds does not reach $85\%$). (c) Ordered eigenvalues of the Hermitian part of the mean combiner-output operator for the MMSE combiner, normalized per geometry by $\tr\bar B_0/4$ and averaged over eight geometries ($400$ fields per geometry). The dash-dot line marks $\lambda/4$.}
\label{fig:digits}
\end{figure*}

Exact restoration matches the ideal backward product ($95.54\%$ for both). Residuals up to $0.2\lambda$ cost little (median $137.5$ to $150$ updates to $85\%$), $0.25\lambda$ and $0.30\lambda$ cost $1.3$ and $2$ times as many updates, and the uncompensated UE ($r=0.47\lambda$, $\R\phi_m=-0.39$) needs a median of $1338$ updates and ends at $82.8\%$. Two observations differ from the scalar model: the slope of the time to target against $\R\phi_m$ is only $-0.45$, and the fixed antenna learns despite a negative scalar correlation. Both outcomes contradict predictions fixed before training, which expected an inverse scaling and a fixed-antenna accuracy of at most $30\%$. The mean cosine between the backpropagated and the exact BS gradients of the fixed antenna is $+0.18$. A receiver-aware trace correlation $\R\rho_{\rm eff}$, computed after the runs from the MMSE combiner, takes the values $1$, $0.98$, $0.92$, $0.82$, $0.69$, $0.53$, and $0.36$ for $r=0$ to $0.30\lambda$ and $-0.17$ at $0.47\lambda$. Against it the slope steepens to $-0.68$. The eigenvalues in Fig.~\ref{fig:digits}(c) agree with the array analysis of Section~\ref{sec:array}. Up to $r=0.25\lambda$ all four eigenvalues of $\Hm(\bar B_r)$ are positive in every geometry, at $0.30\lambda$ the smallest is near zero, and at $0.47\lambda$ only one of four is positive while the trace is negative. Removing the component of the combiner-output gradient in that positive eigenspace before the backward transmission reduces the final accuracy of the fixed antenna from $82.8\%$ to $32.7\%$, whereas removing a random subspace of the same dimension costs $1.2$ points (eight seeds each). Without intervention, the share of the combiner-output gradient energy in that eigenspace grows by $17.9$ percentage points from the first to the last $200$ updates, below the $20$ points predicted before the run. The encoder therefore moves toward the favorable directions only in part, and the removal test, not this drift, carries the causal evidence. The CIFAR-10 intervention of Table~\ref{tab:intervention} adds energy-matched controls to this dimension-matched one.

\emph{Partial compensation in time.} To test Corollary~\ref{cor:partial} in training rather than in the proxy, we repeated the study with the exchange period set by the mechanical cycle ($v=1$~m/s, $V=2.5$~m/s, $28$~GHz, $\tau=5$~ms), the UE advancing $v$ times the cycle period per update, and the ray field redrawn every $1.875$~s of modeled time. With $a=400$~m/s$^2$, the modeled time to $85\%$ accuracy (median over eight seeds) is $2.80$~s for exact compensation and $2.27$~s with a residual of $0.15\lambda$, a $19\%$ saving. With $a=80$~m/s$^2$ it is $8.94$~s and $6.16$~s with a residual of $0.2\lambda$, a $31\%$ saving. Without compensation the period is $\tau=5$~ms and the time to $85\%$ is $5.88$~s, with one of eight seeds not reaching the target. Table~\ref{tab:digitstime} lists all tested points. With $a=400$~m/s$^2$ the time rises again to $3.41$~s at $0.30\lambda$, so the optimum lies inside the tested range. In this network, a residual well inside $\lambda/4$ costs few updates, so the best tested point lies further from exact compensation than the local analysis alone suggests.

\begin{table}[t]\centering\caption{Digit split network with the exchange period set by the mechanical cycle ($1$~m/s, $28$~GHz, $\tau=5$~ms, $V=2.5$~m/s): period, median updates to $85\%$, and modeled time to $85\%$ (eight seeds).}\label{tab:digitstime}
\begin{tabular}{ccccc}\toprule
$a$ (m/s$^2$) & $r/\lambda$ & period (ms) & $K_{85}$ & time (s)\\\midrule
$400$ & $0$ & $18.7$ & $150$ & $2.80$\\
$400$ & $0.15$ & $15.1$ & $150$ & $2.27$\\
$400$ & $0.20$ & $13.9$ & $175$ & $2.44$\\
$400$ & $0.30$ & $11.4$ & $300$ & $3.41$\\
$80$ & $0$ & $59.6$ & $150$ & $8.94$\\
$80$ & $0.15$ & $43.4$ & $150$ & $6.51$\\
$80$ & $0.20$ & $37.9$ & $162.5$ & $6.16$\\
none & $0.47$ & $5.0$ & $1175$ & $5.88$\\\bottomrule
\end{tabular}\end{table}

\section{Beam-Resolved Aging}\label{app:beams}
\begin{figure}[t]\centering
\includegraphics[width=\columnwidth]{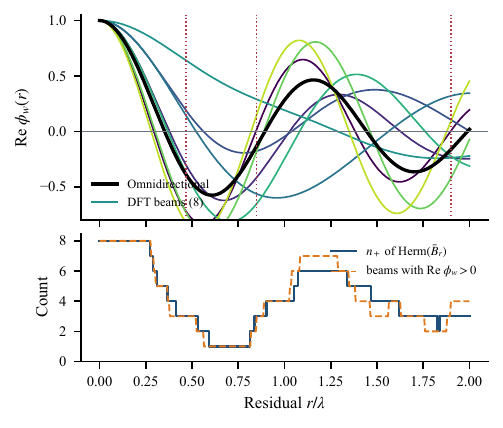}
\caption{Beam-resolved aging on CDL-A with an eight-element DFT combiner. Top: real part of each beam's correlation and of the omnidirectional correlation. Dotted lines mark $0.47\lambda$, $0.85\lambda$, and $1.9\lambda$. Bottom: number of positive eigenvalues of $\Hm(\bar B_r)$ and number of beams with a positive real correlation.}
\label{fig:beams}
\end{figure}

Fig.~\ref{fig:beams} evaluates Lemma~\ref{lem:beam} on CDL-A with an eight-element discrete Fourier transform combiner whose array is perpendicular to the motion. Beams that collect paths nearly perpendicular to the motion keep a positive real correlation well beyond $\lambda/4$, while the omnidirectional correlation turns negative. The number of positive eigenvalues of $\Hm(\bar B_r)$ follows the number of beams with $\R\phi_{\boldsymbol\psi}>0$. Over all eight geometries of Appendix~\ref{app:digits} with an eight-beam DFT combiner, the two counts are equal at $r=0$, $0.25\lambda$, $0.47\lambda$, and $0.85\lambda$ in every geometry ($8$, $8$, $3$, and $3$). Over $r\in[0,2]\lambda$ they agree at $61\%$ of the grid points and differ by one elsewhere. The RMS direction cosine of a single beam ranges from $0.33$ to $0.90$, so the second radius of Lemma~\ref{prop:cert} grows from $0.30\lambda$ for the omnidirectional spectrum to $0.68\lambda$ for the best beam.

\section{Path-Mode Precoding Study}\label{app:pathmode}
This study implements the path-mode transform of Theorem~\ref{thm:paths}(i) and checks its noiseless alignment bound. Joint training without the power condition of part~(iii) is an empirical demonstration, not an instance of the convergence statement. Predictions were fixed before any run with $r>0$, and all $90$ runs are reported. The link has $L=3$ paths at $28$~GHz, a four-element half-wavelength array at the BS, and a four-element array at the UE along its motion. Path gains are i.i.d.\ $\mathcal{CN}(0,1/L)$, and the position is redrawn every update, so the residual is the only source of aging. The UE-side direction cosines along the motion are $(0.10,-0.35,0.45)$ in geometry S, for which $\min_l\cos(2\pi c_lr/\lambda)$ stays positive up to $0.47\lambda$, and $(0.10,-0.35,0.95)$ in geometry P, in which the third path reverses at $0.47\lambda$. The BS-side direction cosines $(-0.55,0.05,0.62)$ give a condition number of $1.34$ for the transmit path-response matrix. The transmitter applies a linear layer to fixed features, the $64$ standardized pixels of the digit dataset and a bias ($1400$ training and $397$ test images). The path-mode encoder is $s=FCx$ with the dual-basis precoder $F=V(V^HV)^{-1}$, and the unstructured encoder is $s=Wx$ with $W\in\mathbb C^{4\times65}$. The UE applies the MMSE combiner of the current forward channel followed by a two-layer network trained with its exact local gradient. Both sides use SGD with step size $0.05$, batch size $32$, and $12000$ updates without power normalization, so that $s=FCx$ holds exactly. The forward and backward noise standard deviations are $0.1$ and $0.05$. Every ten updates we record the alignment $\R\langle g,\hat g\rangle/\norm{g}^2$ of the noiseless backpropagated gradient with the exact one.

Table~\ref{tab:pathmode} lists means over five seeds. With path-mode precoding the alignment never falls below $\min_l\R\zeta_l$, and every residual certified by the pathwise bound, including $0.47\lambda$, which lies beyond the generic condition of part~(iii), ends within the seed spread of restoration. The alignment of the unstructured encoder can fall below the path-mode bound ($0.22$ against $0.24$), as Theorem~\ref{thm:paths} allows, but its accuracy is the same on this geometry. In geometry P at $0.47\lambda$, outside the theorem, both encoders reach $88\%$, below the $94\%$ of restoration. A third geometry with closely spaced departures (condition number $40.5$) was excluded before the runs: at $r=0$ the precoder raised the mean symbol energy to $2420$ against $47$ for the unstructured encoder, the power cost that part~(iii) assumes away. In the reported geometries the mean symbol energy is $61$ to $104$. Path-mode precoding reaches $90\%$ after $500$ updates at every residual, including $r=0$, so this task does not resolve the rate dependence of Theorem~\ref{thm:main}.

\begin{table}[t]\centering\caption{Path-mode precoding study (means over five seeds): final accuracy (\%) and recorded alignment, smallest/mean.}\label{tab:pathmode}
\setlength{\tabcolsep}{4pt}\begin{tabular}{ccc|cc|cc}\toprule
 & & & \multicolumn{2}{c|}{path-mode} & \multicolumn{2}{c}{unstructured}\\
geom. & $r/\lambda$ & $\min_l\R\zeta_l$ & acc. & align. & acc. & align.\\\midrule
S & $0$ & $1.00$ & $93.5$ & $1.00/1.00$ & $94.3$ & $1.00/1.00$\\
S & $0.1$ & $0.96$ & $94.1$ & $0.96/0.98$ & $94.2$ & $0.93/0.98$\\
S & $0.2$ & $0.84$ & $94.2$ & $0.85/0.91$ & $93.5$ & $0.80/0.91$\\
S & $0.3$ & $0.66$ & $94.1$ & $0.67/0.82$ & $93.0$ & $0.62/0.81$\\
S & $0.4$ & $0.43$ & $93.6$ & $0.45/0.69$ & $93.8$ & $0.37/0.67$\\
S & $0.47$ & $0.24$ & $94.3$ & $0.27/0.59$ & $94.0$ & $0.22/0.58$\\
P & $0$ & $1.00$ & $93.8$ & $1.00/1.00$ & $94.3$ & $1.00/1.00$\\
P & $0.2$ & $0.37$ & $93.7$ & $0.43/0.78$ & $93.6$ & $0.34/0.75$\\
P & $0.47$ & $-0.94$ & $87.8$ & $0.05/0.70$ & $87.8$ & $-0.61/0.70$\\
\bottomrule\end{tabular}\end{table}

\section{Checks of the Scalar Model, Calibration, and the Adaptive Criterion}\label{app:checks}
Three checks were specified with decision thresholds before any computation. All three use the CDL ray model of the CIFAR-10 study.

\begin{figure*}[t]\centering
\includegraphics[width=\textwidth]{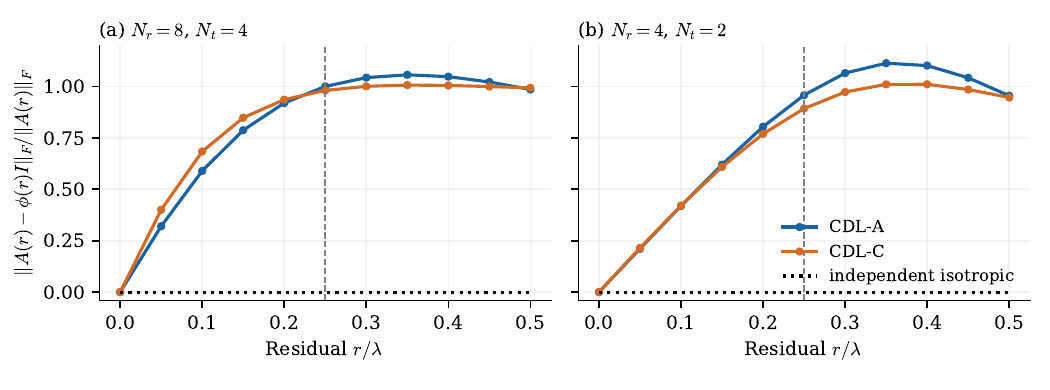}
\caption{Relative deviation $\norm{A(r)-\phi(r)I}_F/\norm{A(r)}_F$ of the linear conditional mean from the scalar model for CDL-A and CDL-C, with the independent isotropic control (bands over three coupling seeds). The dashed line marks $\lambda/4$.}
\label{fig:checkC1}
\end{figure*}

\emph{Scalar conditional mean.} For CDL-A and CDL-C with $8\times4$ and $4\times2$ arrays we computed, from the exact second moments of the ray model, the linear conditional mean $A(r)$ of $\operatorname{vec}H_b$ given $\operatorname{vec}H_f$ and its relative deviation from $\phi(r)I$ (Fig.~\ref{fig:checkC1}). The deviation is about $0.3$ at $r=0.05\lambda$ and about $1$ at $r=0.25\lambda$, so Assumption~\ref{ass:scalar} does not hold for finite ray clusters observed through an array. A control in which every element fades independently with an isotropic field gives $A(r)=\phi(r)I$ to numerical precision. At the gradient level, the predicted mean backward product keeps a cosine of $0.81$ to $0.88$ with $\phi(r)^*H_f^H\delta$ at $0.15\lambda$ and $0.49$ to $0.68$ at $0.25\lambda$. The linearly predicted gradient is also $1.1$ to $2.8$ times longer than the scalar prediction. A regression on fixed cubic features of the forward channel has a held-out error $1.5\%$ to $2.5\%$ higher than the linear predictor, so this feature family detects no nonlinear gain. The array analysis of Section~\ref{sec:array} does not use Assumption~\ref{ass:scalar}.

\begin{figure*}[t]\centering
\includegraphics[width=\textwidth]{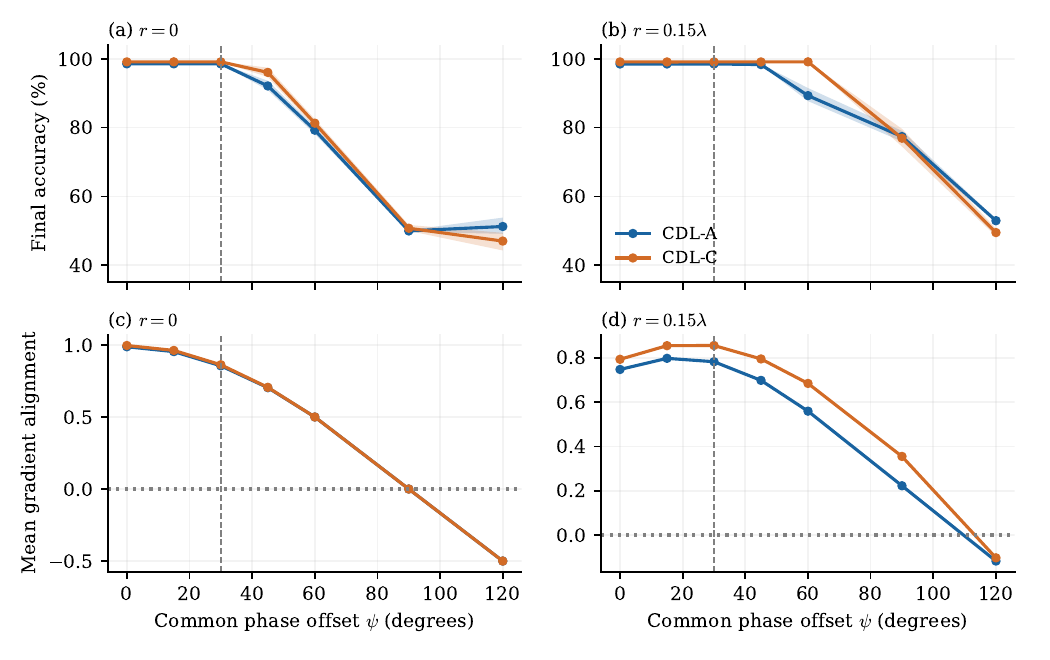}
\caption{Synthetic task with a calibration phase offset between the receiving and transmitting elements: final accuracy and mean gradient alignment versus the offset, at $r=0$ and $r=0.15\lambda$ (eight seeds, $95\%$ bootstrap intervals). The dashed line marks $30^\circ$.}
\label{fig:checkC2}
\end{figure*}

\emph{Calibration phase.} In the synthetic task, a common phase offset $\psi$ between the receiving and transmitting elements leaves accuracy unchanged up to $30^\circ$ ($98.56\%$ on CDL-A and $99.15\%$ on CDL-C at $r=0$), reduces it by $6.4$ and $3.1$ percentage points at $45^\circ$, and reaches chance at $90^\circ$ (Fig.~\ref{fig:checkC2}). Independent Gaussian per-element offsets with a standard deviation of $20^\circ$ cost less than $0.1$ points at $r=0$. At $r=0.15\lambda$ the per-element offsets interact with the residual: $20^\circ$ and $40^\circ$ cost $1.9$ and $10.4$ points on CDL-A and $1.1$ and $3.4$ points on CDL-C, whereas a common offset up to $45^\circ$ costs at most $0.2$ points at this residual. The pre-specified robustness threshold (at most $2$ points at $45^\circ$) was not met, so phase calibration within about $30^\circ$ is a design requirement for discrete positioning.

\begin{figure}[t]\centering
\includegraphics[width=\columnwidth]{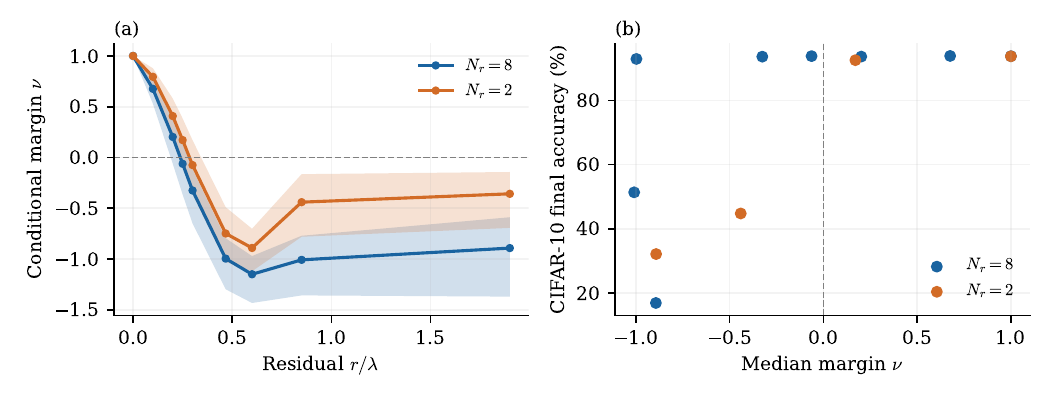}
\caption{Conditional alignment margin of the eight- and two-antenna MMSE UE (median and $10$ to $90\%$ quantiles over fields) and its relation to the CIFAR-10 final accuracy.}
\label{fig:checkC3}
\end{figure}

\emph{Conditional criterion for MMSE processing.} Using the linear conditional mean above, we computed for each forward field the largest $\nu$ for which $\Hm(\E[B_r\mid H_f])-\nu B_0\succeq0$ on the range of $B_0$ (Fig.~\ref{fig:checkC3}). For the eight-antenna MMSE UE the median margin is negative from $r=0.25\lambda$ on, and at $0.47\lambda$ it is negative for every field, whereas this configuration trains to $92.9\%$. The criterion of Proposition~\ref{prop:policy} guarantees alignment for every adaptive direction, and the realized MMSE gradients do not use the unfavorable directions, consistent with the intervention of Table~\ref{tab:intervention}. The criterion is therefore sufficient but conservative for this receiver. Because the true conditional mean may differ from its linear approximation, this check does not provide a strict counterexample to the conditional guarantee.

\section{Complete CIFAR-10 Results}\label{app:cifar}
This appendix reports every CIFAR-10 configuration behind Section~\ref{sec:eval}. All runs use the setup of Section~\ref{sec:eval-cifar} and the same three seeds, and no run was repeated or removed. Each block of configurations was specified, with its predictions, before it was run.

\begin{table*}[t]\centering\caption{CIFAR-10 at $1$~m/s with MMSE combining: final test accuracy, mean (range) over three seeds, accuracy after $10$ epochs, and epochs to $90\%$ for each seed. The last two columns give the number of positive eigenvalues and the largest eigenvalue of $\Hm(\bar B_r)$, normalized by $\tr\bar B_0/N_t$ and computed on the three training geometries before training.}\label{tab:cifarfull}
\begin{tabular}{llccccc}\toprule
UE & backward signal & final acc. (\%) & epoch $10$ (\%) & epochs to $90\%$ & $n_+$ & $\lambda_{\max}$\\\midrule
$8\times4$ & ideal backward product & $94.02$ ($93.85$--$94.26$) & $84.10$ & $26, 25, 24$ & & \\
 & restoration, $r=0$ & $93.81$ ($93.60$--$94.10$) & $81.56$ & $27, 26, 27$ & $4$ & $1.01$\\
 & $r=0.10\lambda$ & $93.88$ ($93.79$--$94.06$) & $84.53$ & $27, 24, 26$ & $4$ & $0.99$\\
 & $r=0.20\lambda$ & $93.68$ ($93.60$--$93.78$) & $83.50$ & $29, 25, 26$ & $4$ & $0.91$\\
 & $r=0.25\lambda$ & $93.82$ ($93.62$--$94.16$) & $83.53$ & $24, 26, 29$ & $4$ & $0.86$\\
 & $r=0.30\lambda$ & $93.66$ ($93.30$--$94.03$) & $83.33$ & $29, 28, 26$ & $3$ to $4$ & $0.79$\\
 & no compensation, $r=0.47\lambda$ & $92.95$ ($92.56$--$93.17$) & $77.70$ & $31, 31, 30$ & $1$ & $0.54$\\
 & $r=0.85\lambda$ & $51.35$ ($37.51$--$60.03$) & $32.42$ & not reached & $1$ & $0.11$\\
 & $r=1.90\lambda$ & $16.85$ ($4.11$--$30.11$) & $15.86$ & not reached & $1$ & $0.13$\\\midrule
$2\times4$ & ideal backward product & $93.72$ ($93.51$--$93.93$) & $82.39$ & $28, 28, 26$ & & \\
 & restoration, $r=0$ & $93.77$ ($93.64$--$93.88$) & $82.62$ & $27, 26, 25$ & $4$ & $1.53$\\
 & $r=0.25\lambda$ & $92.56$ ($92.45$--$92.65$) & $77.64$ & $32, 32, 34$ & $4$ & $0.82$\\
 & no compensation, $r=0.47\lambda$ & $85.52$ ($82.88$--$87.06$) & $44.82$ & not reached & $1$ & $0.32$\\
 & $r=0.60\lambda$ & $32.10$ ($26.09$--$37.11$) & $21.50$ & not reached & $1$ & $0.07$\\
 & $r=0.85\lambda$ & $44.76$ ($39.03$--$53.32$) & $22.63$ & not reached & $0$ & $-0.03$\\\midrule
$2\times2$ & ideal backward product & $93.74$ ($93.33$--$93.96$) & & & & \\
 & restoration, $r=0$ & $93.92$ ($93.74$--$94.23$) & & & $2$ & $1.03$\\
 & no compensation, $r=0.47\lambda$ & $46.61$ ($43.81$--$51.94$) & & & $1$ & $0.23$\\\bottomrule
\end{tabular}\end{table*}

\begin{table*}[t]\centering\caption{CIFAR-10 at higher speeds, with discrete positioning, and with the CSI-free UE: final test accuracy, mean (range) over three seeds. The trajectory advance per update scales with the speed. Discrete positioning uses half-wavelength-spaced elements along the motion, and its residual is given in parentheses.}\label{tab:cifarspeed}
\begin{tabular}{lllcccc}\toprule
UE & receiver & speed & ideal backward & restoration & discrete positioning & no compensation\\\midrule
$8\times4$ & MMSE & $1$~m/s & $94.02$ & $93.81$ & $93.92$ ($93.79$--$94.00$), $r=-0.03\lambda$ & $92.95$\\
$8\times4$ & MMSE & $1.82$~m/s & $93.84$ ($93.75$--$93.91$) & $93.84$ ($93.65$--$93.99$) & $93.54$ ($93.29$--$93.84$), $r=-0.15\lambda$ & $58.81$ ($50.69$--$64.71$)\\
$8\times4$ & MMSE & $4.07$~m/s & $93.78$ ($93.71$--$93.82$) & $93.88$ ($93.75$--$93.98$) & $93.96$ ($93.90$--$94.05$), $r=-0.10\lambda$ & $16.07$ ($9.95$--$20.91$)\\
$2\times4$ & MMSE & $1$~m/s & $93.72$ & $93.77$ & $93.69$ ($93.37$--$93.87$), $r=-0.03\lambda$ & $85.52$\\
$2\times4$ & MMSE & $1.82$~m/s & $93.93$ ($93.80$--$94.15$) & $93.79$ ($93.58$--$93.93$) & $93.64$ ($93.44$--$94.00$), $r=-0.15\lambda$ & $53.21$ ($50.81$--$57.53$)\\
$2\times2$ & MMSE & $1.82$~m/s & & $93.83$ ($93.76$--$93.96$) & & $31.07$ ($29.00$--$33.36$)\\
$8\times4$ & CSI-free & $1$~m/s & $93.72$ ($93.56$--$93.89$) & $93.69$ ($93.49$--$93.81$) & & $10.00$ (all seeds)\\
$2\times2$ & CSI-free & $1$~m/s & $93.59$ ($93.40$--$93.76$) & $93.78$ ($93.60$--$93.88$) & & $10.00$ (all seeds)\\\bottomrule
\end{tabular}\end{table*}

\emph{Eight-antenna UE.} The main sweep (Table~\ref{tab:cifarfull}, top) carried three predictions. Restoration came within $0.21$ points of the ideal backward product, as predicted. The prediction that the uncompensated UE would lose at least $5$ points or need three times as many epochs to $80\%$ failed: it loses $0.86$ points and needs $1.5$ times as many epochs. The predicted rank correlation of at most $-0.8$ between the residual and the final accuracy was not met ($-0.71$). The remaining blocks were specified after this outcome and test the array explanation of Section~\ref{sec:array}.

\emph{Small arrays and large residuals.} With two receive antennas, exact restoration matches the ideal backward product within $0.05$ points, and the accuracy over the tested residuals has a rank correlation of $0.90$ with $\lambda_{\max}$. The ordering is not strict at the low end. At $0.60\lambda$, where one weak positive eigenvalue remains, the accuracy is lower than at $0.85\lambda$, where none remains. The eight-antenna UE also fails once the residual reaches $0.85\lambda$ or $1.9\lambda$, although $\lambda_{\max}$ stays positive there. A weakly positive eigenvalue is thus not enough for learning within the epoch budget. The $2\times2$ link fails without compensation at $1$ and $1.82$~m/s although it carries no more streams than receive antennas.

\emph{Speeds and discrete positioning.} With the trajectory advanced in proportion to the speed, restoration stays within $0.21$ points of the ideal backward product at every tested speed and array, while the uncompensated accuracy drops to between $16\%$ and $59\%$ (Table~\ref{tab:cifarspeed}). Discrete positioning stays within $0.3$ points of exact restoration at the same speed in every tested configuration.

\emph{CSI-free UE.} The CSI-free UE matches its ideal backward product within $0.2$ points under restoration and stays at chance without compensation in every seed, for both the $8\times4$ and the $2\times2$ link. Its residual sweep (Fig.~\ref{fig:csifree}) gives $93.63\%$, $93.44\%$, $92.49\%$, and $16.12\%$ at $r=0.10$, $0.20$, $0.25$, and $0.30\lambda$, and its accuracy ranks the residuals exactly as the real part of the scalar correlation does.

\section{Synthetic-Task Details}\label{app:synthetic}
\emph{Complete residual sweep.} Table~\ref{tab:ma1} lists every configuration of the CDL-A residual study in Fig.~\ref{fig:residual}. The compensation ratio $q/(v\tau)$ sets the residual $r=(1-q/(v\tau))v\tau$, with $v\tau=0.389\lambda$, $1.30\lambda$, and $3.89\lambda$ at $3$, $10$, and $30$~km/h. Configurations with the same residual give the same result, because the trained system depends on the motion only through $r/\lambda$. The ideal backward product reaches $98.47\%$. The rank correlation of $0.953$ between $\R\phi_m$ and the accuracy has a seed-bootstrap interval of $[0.842,0.971]$. The second-lobe point at $1.17\lambda$ exceeds the first-lobe point of similar correlation magnitude by $30.7$ points, with interval $[26.9,35.0]$. The fixed antenna at $10$~km/h also lies in the second positive lobe ($r=1.30\lambda$, $\R\phi_m=0.33$) and reaches $84.7\%$. The first zero of $\R\phi_m$ lies at $0.352\lambda$. At $30$~km/h the accuracy first falls below $80\%$ at a ratio of $0.9$, next to the ratio $0.91$ at which the residual reaches this zero. At $10$~km/h it falls below $80\%$ at a ratio of $0.8$, where $\R\phi_m$ is still $0.37$, so there the loss precedes the zero within the $120$-update budget.

An independent replication with eight new seeds gives $98.47\%$ at the second-lobe point, $66.1\%$ and $50.9\%$ at ratios $0.8$ and $0.6$ for $10$~km/h, $98.33\%$ at a ratio of $0.6$ for $3$~km/h, and $48.5\%$ and $84.4\%$ for the fixed antenna at $3$ and $10$~km/h.

\begin{table*}[t]\centering\caption{Complete CDL-A residual sweep (synthetic task, eight seeds): residual, real part of the correlation, and mean test accuracy (\%) for each speed and compensation ratio. A ratio of $0$ is the fixed antenna.}\label{tab:ma1}
\begin{tabular}{c|ccc|ccc|ccc}\toprule
 & \multicolumn{3}{c|}{$3$~km/h} & \multicolumn{3}{c|}{$10$~km/h} & \multicolumn{3}{c}{$30$~km/h}\\
$q/(v\tau)$ & $r/\lambda$ & $\R\phi_m$ & acc. & $r/\lambda$ & $\R\phi_m$ & acc. & $r/\lambda$ & $\R\phi_m$ & acc.\\\midrule
$0$ & $0.389$ & $-0.142$ & $49.01$ & $1.297$ & $0.327$ & $84.73$ & $3.892$ & $-0.164$ & $51.44$\\
$0.5$ & $0.195$ & $0.622$ & $85.52$ & $0.649$ & $-0.560$ & $50.46$ & $1.946$ & $-0.082$ & $51.75$\\
$0.6$ & $0.156$ & $0.750$ & $98.33$ & $0.519$ & $-0.500$ & $48.82$ & $1.557$ & $-0.235$ & $45.66$\\
$0.7$ & $0.117$ & $0.856$ & $98.36$ & $0.389$ & $-0.142$ & $49.01$ & $1.167$ & $0.463$ & $98.35$\\
$0.8$ & $0.078$ & $0.935$ & $98.42$ & $0.259$ & $0.375$ & $67.66$ & $0.778$ & $-0.332$ & $50.28$\\
$0.85$ & $0.058$ & $0.963$ & $98.43$ & $0.195$ & $0.622$ & $85.52$ & $0.584$ & $-0.571$ & $48.56$\\
$0.9$ & $0.039$ & $0.984$ & $98.44$ & $0.130$ & $0.823$ & $98.36$ & $0.389$ & $-0.142$ & $49.01$\\
$0.95$ & $0.019$ & $0.996$ & $98.45$ & $0.065$ & $0.954$ & $98.42$ & $0.195$ & $0.622$ & $85.52$\\
$1$ & $0$ & $1$ & $98.43$ & $0$ & $1$ & $98.43$ & $0$ & $1$ & $98.43$\\\bottomrule
\end{tabular}\end{table*}

\emph{Matched comparison.} Table~\ref{tab:benchfull} gives the values behind Fig.~\ref{fig:benchsupp}. Restoration equals the ideal backward product within $0.01$ points. Polar correction loses at most $0.53$ points, and delayed feedback loses up to $7.9$ points on CDL-A at $30$~km/h.

\begin{table}[t]\centering\caption{Matched comparison (synthetic task, eight seeds): mean test accuracy (\%) and, in parentheses, paired cross-entropy excess over the ideal backward product. The ideal backward product reaches $98.55\%$ on CDL-A and $99.14\%$ on CDL-C.}\label{tab:benchfull}
\setlength{\tabcolsep}{3pt}\begin{tabular}{lcccc}\toprule
 & restoration & polar & delayed & fixed\\\midrule
A, $3$~km/h & $98.56$ ($0$) & $98.55$ ($0.001$) & $93.55$ ($0.101$) & $48.74$ ($0.638$)\\
A, $30$~km/h & $98.56$ ($0$) & $98.46$ ($0.004$) & $90.64$ ($0.172$) & $48.81$ ($0.669$)\\
C, $3$~km/h & $99.14$ ($0$) & $99.07$ ($0.003$) & $98.79$ ($0.014$) & $49.12$ ($0.655$)\\
C, $30$~km/h & $99.14$ ($0$) & $98.61$ ($0.021$) & $96.82$ ($0.065$) & $48.93$ ($0.665$)\\\bottomrule
\end{tabular}\end{table}

\emph{Positioning errors.} Table~\ref{tab:biasfull} completes Fig.~\ref{fig:bias}. Besides the accuracy it lists, along the training trajectory, the mean cosine between the backpropagated and the exact encoder gradients and the fraction of updates in which their inner product is negative. A bias of $0.383\lambda$ makes this fraction $47.9\%$, close to that of a random direction, whereas Gaussian jitter of the same RMS displacement keeps it at $8.4\%$. A second run lets all four elements share one jitter draw per exchange, as for elements on a common rail. At $\sigma=0.15\lambda$, $0.383\lambda$, and $0.45\lambda$ this common jitter gives $98.46\%$, $98.49\%$, and $98.42\%$, within $0.1$ points of independent per-element jitter in the same run, and the bias of $0.383\lambda$ gives $53.3\%$ there.

\begin{table}[t]\centering\caption{Fixed bias and independent Gaussian jitter in the planar isotropic field (synthetic task, eight seeds). The fixed antenna corresponds to a bias of $0.5\lambda$, and the ideal backward product reaches $98.59\%$.}\label{tab:biasfull}
\begin{tabular}{lcccc}\toprule
error & size & acc. (\%) & mean cosine & negative updates\\\midrule
none & $0$ & $98.51$ & $0.883$ & $0\%$\\
bias & $0.15\lambda$ & $98.51$ & $0.794$ & $0.2\%$\\
bias & $0.383\lambda$ & $50.32$ & $0.013$ & $47.9\%$\\
bias & $0.5\lambda$ & $43.24$ & $-0.457$ & $90.1\%$\\
jitter & $0.15\lambda$ & $98.57$ & $0.803$ & $0.1\%$\\
jitter & $0.383\lambda$ & $98.42$ & $0.471$ & $8.4\%$\\
jitter & $0.45\lambda$ & $98.38$ & $0.397$ & $12.7\%$\\\bottomrule
\end{tabular}\end{table}